%% file: dr14_lrg_bao_arxiv.tex
\documentclass{aastex6}

\newcommand{\hmpc}{$h^{-1}$~Mpc}

\newcommand{\aiso}{\alpha_{\rm iso}}
\newcommand{\aperp}{\alpha_\perp}
\newcommand{\apar}{\alpha_\parallel}

\usepackage{rotating}

\begin{document}


\title{The SDSS-IV extended Baryon Oscillation 
Spectroscopic Survey:  
Baryon Acoustic Oscillations at redshift of 0.72 with the DR14 Luminous Red Galaxy Sample}



\input{authors}


\begin{abstract}

  The extended Baryon Oscillation Spectroscopic Survey (eBOSS) Data
  Release 14 sample includes 80,118 Luminous Red
  Galaxies. By combining these galaxies with the high-redshift tail of
  the BOSS galaxy sample, we form a sample of LRGs  at an effective
  redshift $z=0.72$, covering an effective volume of
  0.9~Gpc$^3$. 
  We introduce new techniques to account for spurious fluctuations
  caused by targeting and by redshift failures which were validated 
  on a set of mock catalogs.   
  This analysis is sufficient to provide a $2.6$\% measurement of spherically 
  averaged BAO, $D_V(z=0.72) = 2353^{+63}_{-61} (r_d/r_{d,\rm{fid}})$\hmpc, 
  at 2.8$\sigma$ of significance.
  Together with the recent quasar-based BAO measurement
  at $z=1.5$, and forthcoming Emission Line Galaxy-based measurements,
  this measurement demonstrates that eBOSS is fulfilling 
  its remit of extending the range of
  redshifts covered by such measurements, laying the ground work for
  forthcoming surveys such as the Dark Energy Spectroscopic Survey 
  and Euclid.

\end{abstract}




\section{Introduction} \label{sec:intro}
Over the last decade, the expansion history of the Universe has been
measured to percent-level precision using observations of the baryon
acoustic oscillations (BAO, \citealt{PeeblesYu70, SZ70, BondEfstathiou87}) 
in the distribution of galaxies. Multiple
measurements from an increasing number of surveys have provided robust
support for the standard $\Lambda$CDM cosmological model. Key
early surveys such as the Sloan Digital Sky Survey (SDSS;
\citealt{York00}) and the 2-degree Field Galaxy Redshift Survey
(2dFGRS; \citealt{Colless03}) generated spectroscopic samples for the
BAO measurements given in \citet{Eis05, Per07} and 
\citet{Percival01, Cole05}, respectively. These spectroscopic programs 
were followed by the WiggleZ survey at higher
redshift \citep{Wigglez} and 
the 6-degree Field Galaxy Survey (6dFGS;
\citealt{6dF}) at lower redshift, also measuring  BAO 
(\citealt{Blake11, Beutler11}, respectively). The Baryon
Oscillation Spectroscopic Survey (BOSS; \citealt{Dawson13}), conducted
as part of the Sloan Digital Sky Survey III (SDSS-III;
\citealt{Eis11}), provided the first BAO measurements with precision 
better than 1\%
\citep{Anderson2012,Anderson2014a,Anderson2014b,TojeiroEtAl14}. 
Results from the final sample of BOSS galaxies were presented in
\citet{Alam17}, while results using the final sample of BOSS 
Lyman-$\alpha$ forests are presented in \citet{Bautista17} and 
\citet{Bourboux17}.

These BAO measurements are all broadly consistent with a
$\Lambda$CDM cosmological model as inferred from  Planck satellite
measurements \citep{Planck16I, Planck16XIII}. 
Even so, the low value of the cosmological
constant constrained by these data has yet no compelling theoretical 
explanation (see \citealt{Weinberg13} for a review). 
The increasing precision of cosmological measurements has 
renewed the interest in alternative models that predict similar behavior
with a different mechanism causing cosmological acceleration
(see \citealt{Clifton12} for a review). 
The combination of BAO measurements with measurements from Redshift-Space
Distortions, supernovae, weak-lensing and other low-redshift
cosmological probes has therefore recently seen renewed focus, with
many planned upcoming experiments.
The Taipan survey will observe
approximately 2 million galaxies over half the sky at $z<0.4$ 
\citep{Taipan17}. At higher redshifts, the Dark Energy Spectroscopic 
Instrument (DESI; \citealt{DESII, DESIII}) 
and European Space Agency Euclid mission
\citep{Euclid} will observe an order of magnitude
more galaxies. The extended BOSS (eBOSS; \citealt{Dawson16}), part of
the SDSS-IV experiment \citep{Blanton17} is the largest spectroscopic
galaxy survey running at this time.
eBOSS extends the redshift range beyond the BOSS galaxy sample, 
to redshifts that will be covered by DESI and Euclid.

In this paper we present a BAO measurement at an effective
redshift $z=0.72$ using Luminous Red Galaxies (LRGs) observed by
eBOSS. This sample, described in
Section~\ref{sec:data}, is designed to extend the CMASS sample from
BOSS \citep{Reid16} to higher redshift. This sample of galaxies is
supplemented by the final BOSS galaxy sample. In addition to the new
data, the analysis methods in this work were improved from previous BOSS
studies:
\begin{itemize}
\item spurious fluctuations caused by non-cosmological variations in 
target density  are modeled  from 
multiple linear regression; 
\item enhanced characterization of ``redshift failures'', i.e., 
spectra where the redshift is not measured with sufficient 
statistical significance; 
\item  corrections for targeting inhomogeneity and spectroscopic 
incompleteness are applied on the random catalog instead of
  up-weighting galaxies, thus reducing shot-noise and 
   systematic errors in the two dimensional correlation function.
\end{itemize} 

The structure of the paper is as follows. Section~\ref{sec:data}
describes our data and mock catalogs. Our new treatment of photometric
systematic errors and redshift failures is presented in
section~\ref{sec:syst_photo} and \ref{sec:redshiftfail}, and tested
using mock catalogs. 
The model for BAO in the correlation function and the technique for 
reconstructing integrated bulk flows that degrade the BAO feature 
are presented in section~\ref{sec:methods}. 
Section~\ref{sec:results} shows results on data
and the main BAO measurement. Table~\ref{tab:cosmologies} presents the
cosmological models employed in our work.

\begin{table}[t]
\centering
\caption{Cosmological parameters used in this work.}
\label{tab:cosmologies}
\begin{tabular}{ccc}
\hline
\hline
 & Fiducial & Mocks \\
\hline
$\Omega_M h^2$ & 0.1417 & 0.1421 \\
$= \Omega_c h^2$ & 0.1190  & 0.1196 \\
$+ \Omega_b h^2$ & 0.0220 & 0.0225 \\
$+ \Omega_\nu h^2$ & 0.0006 & 0 \\
$h$ & 0.676 & 0.7 \\
$N_\nu$ & 3  & 3 \\
$\sigma_8$ & 0.8 & 0.816 \\
$n_s$ & 0.97 & 0.97 \\
\hline
$r_d$ [Mpc] & 147.78 & 147.13 \\
$D_H(z=0.71)/r_d$ & 20.06 & 19.82 \\
$D_M(z=0.71)/r_d$ & 17.68 & 17.31 \\
$D_V(z=0.71)/r_d$ & 16.45 & 16.15 \\
\hline
\hline
\end{tabular}
\vspace{0.5cm}
\end{table}

\section{Data} 
\label{sec:data}

The sample of galaxies used in this work is mainly composed of
Luminous Red Galaxies (LRGs) observed spectroscopically  
during the first two years of
the Extended Baryon Oscillation Spectroscopic Survey (eBOSS,
\citealt{Dawson16}) --- the cosmological component of the fourth
generation of the Sloan Digital Sky Survey (SDSS-IV,
\citealt{Blanton17}).  In order to increase tracer density, we combine
the eBOSS sample with the $z>0.6$ BOSS CMASS galaxies \citep{Alam17}
trimmed to the area covered by eBOSS observations.  These BOSS galaxies
represent about a third of the full sample.  
These data are found in the SDSS Data Release 
14\footnote{\url{sdss.org/dr14}} \citep{DR14}.

\subsection{Galaxy sample and redshift estimators}
The eBOSS spectroscopic targets were selected from optical 
(SDSS DR13, \citealt{DR13}) and
infrared (from the Wide-field Infrared Survey Explorer, WISE,
\citealt{Wright10}) imaging data with infrared forced photometry 
applied over positions of SDSS sources \citep{Lang14}.  
The LRG target selection is fully
described in \cite{Prakash16}.
 The selection algorithm was informed 
by \citet{Prakash15} and
applied over the full BOSS imaging footprint, yielding about 60
deg$^{-2}$ LRG targets, of which 50~deg$^{-2}$ were observed 
spectroscopically. In short, the main (extinction corrected) 
SDSS magnitude cuts of this sample are given by
\begin{equation}
z < 19.95 \ \ \ \ 19.9 < i < 21.8 
\end{equation}
which makes the eBOSS LRG galaxies a completely disjoint 
set (in magnitudes, not redshift) from CMASS galaxies \citep{Eis11}.
Additional color cuts,
\begin{equation}
r-i > 0.98 \ \ \ \ i-z > 0.625 \ \ \ \ r-W1 > 2(r-i),
\end{equation}
yield, on average, higher redshift galaxies than CMASS 
while avoiding star contamination.
The selection was tested over 466~deg$^2$ covered during 
the Sloan Extended Quasar, ELG, and LRG Survey (SEQUELS). 
An overview of this pilot survey can be
found in \citet{Dawson16} and in the DR12 data release
\citep{DR12}.

Spectra were obtained by the Sloan 2.5m telescope at Apache Point
Observatory, New Mexico, USA \citep{Gunn06}. Two multi-object
spectrographs \citep{Smee13}  simultaneously project 1000 spectra
per exposure, including about 20 calibration stars and
$\sim 80$ empty regions (for modeling sky subtraction). 
Spectra cover a wavelength range from 3,600 to 10,000~\AA\ 
with a resolution $R = 1500 - 2600$. Sets of 15
minute exposures were taken until a typical target with $g=21.2$
and $i=20.2$ reaches a signal-to-noise ratio of 3.16 per pixel 
in the $g$ band and 4.7 per pixel in the $i$ band.

Spectra were extracted, sky-subtracted, flux-calibrated and co-added
using version \texttt{v5\_10\_0} of the software
\texttt{idlspec2d}\footnote{Available at
\url{sdss.org/dr14/software/products/}}. 
The extraction algorithm has improved since BOSS, its description
can be found in Appendix B of \citet{Bautista17}. 
Recent improvements on
co-addition and flux-calibration are described in \citet{Hutchinson16}
and \citet{Jensen16}.  The final eBOSS LRG spectra were  
classified and their redshifts measured primarily by
\texttt{redmonster}\footnote{\url{https://github.com/timahutchinson/redmonster/}}
\citep{Hutchinson16}, complemented by redshifts obtained using
\texttt{spec1d} \citep{Bolton12}.  On average 10\% of the
eBOSS LRG sample lacks a statistically confident redshift estimate due 
mainly to
the low signal-to-noise ratio of their spectra. In 
section~\ref{sec:redshiftfail} we discuss how to this failure rate 
depends on many characteristics of
the observations, e.g., position of the fiber in the focal plane or
position of the trace in the CCD. In that section we present the 
methods used to account for these
fluctuations in our clustering measurement.

\subsection{Catalog creation}
\label{sec:catalog}

One important step of the clustering analysis is to determine the
survey mask.  Usually the mask is defined by a set of random points
Monte-Carlo sampling the volume covered by the survey, referred as the
``random catalog'' or simply ``the randoms''. This random catalog also
accounts for angular variations in spectroscopic completeness of the
survey.  We follow here a similar procedure to that used 
for the final BOSS galaxy clustering measurements,
described in \citet{Reid16}.

Starting from the photometric target sample, we first veto 
objects in ``bad'' photometric regions of the sky which were included
in the target selection process. We exclude 
regions around stars in the Tycho catalog \citep{Hog00} 
with Tycho $B_T$ magnitudes 
within [6, 11.5] with a magnitude-dependent radius ranging from 
3.4$'$ to 0.8$'$. An additional mask excludes regions 0.1 to 1.5$^{\circ}$ 
in radius around bright galaxies and other objects \citep{Rykoff14}. 
Bright objects in WISE imaging are also masked. Regions 
of radius from 16.6$'$ to 2$'$ are masked around sources 
with W1 magnitudes ranging from 2 to 8. 
We removed regions where galactic extinction $E(B-V) > 0.15$ or where
the seeing is larger than 2.3, 2.1, 2.0$'$ in $g$, $r$ and $i$ bands, 
respectively.
Bad photometric regions and bright objects mask 4.5\% of the targets 
in the NGC and 12.1\% of targets in the SGC. 
Since the eBOSS quasars and several other target classes have 
priority during the fiber assignment procedure \citep{Dawson16},
we are unable to obtain spectra from LRG targets that lie 
less than 62$''$ (corresponding angular diameter of a fiber in the sky) 
from a higher priority target. This results in 7.7\% of LRG targets being
masked by quasar fibers in the NGC and 7.0\% in the SGC. 
We expect that the effect on the LRG clustering
caused by these knockouts is negligible given that other sample
are relatively sparse and have little overlap in redshift with the galaxy 
sample. We leave tests on this assumption to future work.  
We mask few tens of the remaining LRG targets that lie
in the center of the focal plane, where a center post holds the 
plate and prevents fibers from being assigned within the 92$''$
central radius. 
The total masked area
is 12.3\% for the NGC and 18.2\% for the SGC.
Targets that are masked in this process do not account in the 
fiber completeness calculations (see below).

The spectroscopic sample is then matched to the remaining targets.  A
small fraction of targets do not receive an optical fiber and
therefore have no spectra or redshift information.  
Some of these missing redshifts are
caused by the impracticability of placing two optical fibers on LRG
targets closer than 62$''$. We refer to these as fiber collisions. These
collisions impact the small-scale clustering since they preferably
occur in over-dense regions. Different methods to correct for the
effect of collisions have been studied in the past (e.g., \citealt{Guo12,
  Bianchi17}). For our catalogs, we simply up-weight by one the target
with redshift information as performed in \cite{Alam17}. The impact on 
correlations larger than 10\hmpc\ is negligible \citep{Guo12}.  Other
targets do not receive a fiber because of the limited number of fibers
available per observation (1000 fibers). 

We correct for these missing targets
by downsampling the random catalog so it follows the computed
completeness. We define fiber assignment completeness as:
\begin{equation}
C = \frac{N_{\rm gal} + N_{\rm qso} + N_{\rm star} + N_{\rm cp} + 
N_{\rm fail}}{N_{\rm targ}}  
\label{eq:completeness}
\end{equation}
where $N_{\rm gal}$ is the number of confirmed LRGs with redshifts,
$N_{\rm qso}$ and $N_{\rm star}$ are the numbers of quasars and stars
found among LRG targets (incorrect target classes), $N_{\rm cp}$ is the
number of LRG targets without spectra due to a collision with a galaxy
with a redshift (they count as being observed), and $N_{\rm fail}$ is
the number targets with spectra but without a confident redshift (we
correct for redshift failures in a later process, see
Section~\ref{sec:redshiftfail}).  The fiber completeness is computed per
``sector'', where a sector is a connected region of the sky defined by
a unique set of plates.  
We exclude sectors where the fiber completeness is
below 50\% to avoid regions covered by
multiple plates, where unfinished observations potentially introduce
an artificial pattern of clustering. 

\begin{figure}[t]
\centering
\includegraphics[width=0.49\textwidth]{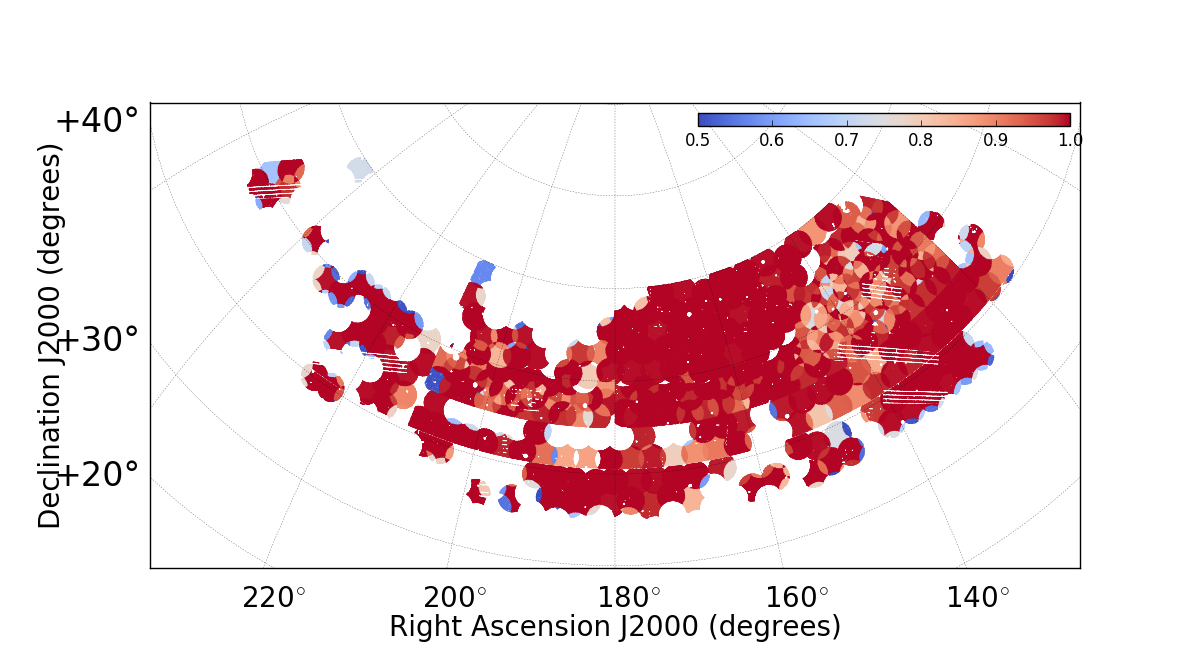}
\includegraphics[width=0.49\textwidth]{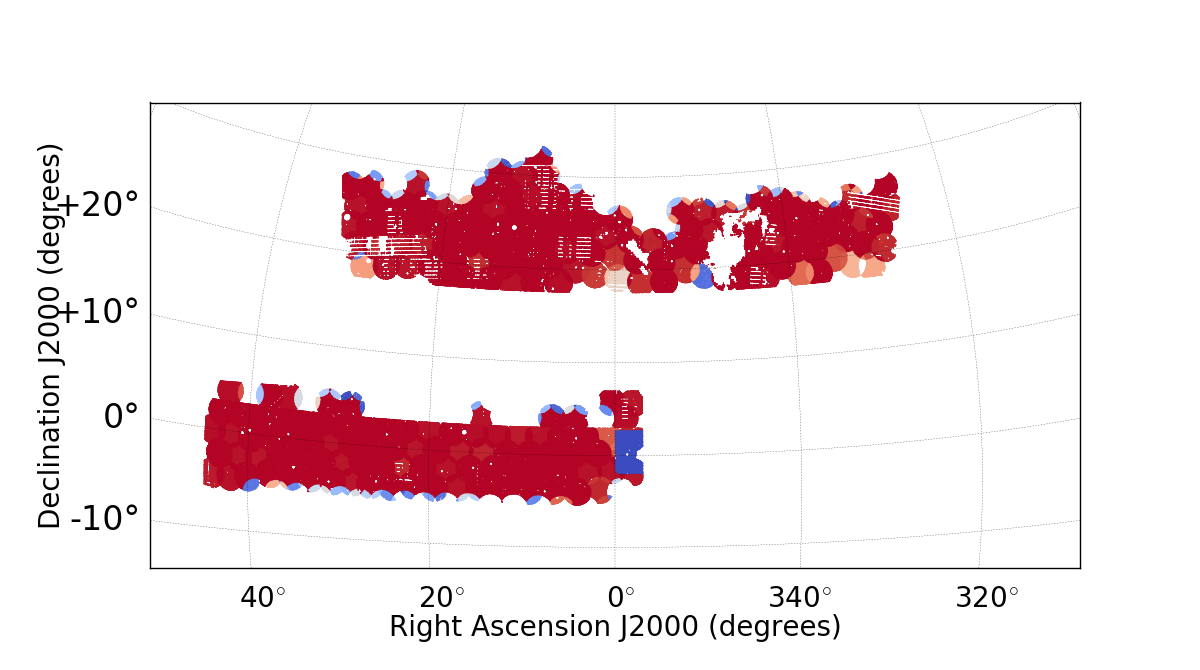}
\caption{The footprint of the DR14 LRG sample. The colors show the 
fiber completeness per region for the eBOSS sample only. 
Regions with fiber completeness below 
0.5 were removed from the final sample.}
\label{fig:footprint}
\end{figure}

Fig.~\ref{fig:footprint} displays the footprint of the DR14 LRG sample 
where the color-coding indicates the corresponding fiber completeness. 
The top of Table~\ref{tab:samples_stats} presents the number of LRGs, 
the total effective area (weighted by completeness) of our sample 
and the effective volume (defined below) of our samples.

\begin{table}[t]
\centering
\caption{Survey specification of the eBOSS LRG sample used in this analysis}
\label{tab:samples_stats}
\begin{tabular}{ccrrrrrrrc}
\hline
\hline
Survey & Cap & $N_{\rm gal}$ & $N_{\rm zfail}$ & $N_{\rm cp}$ & $N_{\rm qso}$ & $N_{\rm star}$ & $A_{\rm eff}$ [deg$^2$] &  $V_{\rm eff} \ [{\rm Gpc}^3]$ \\
     &  NGC & 45826 & 4957 & 2263 & 18 & 2897 & 1033.4 & 0.356 \\
eBOSS & SGC & 34292 & 4366 & 1687 & 18 & 4273 & 811.6 & 0.262 \\
    & \bf{Total} & \bf{80118} & \bf{9323} & \bf{3950} & \bf{36} & \bf{7170} &  \bf{1844.0} & \bf{0.618} \\
\hline
     eBOSS +       & NGC & 71975 & & & & & & 0.511\\
CMASS & SGC & 54582 & & & & & & 0.389 \\
    $(0.6 < z < 1.0)$     & \textbf{Total} & \textbf{126557} & & & & & & \textbf{0.900} \\
\hline
\end{tabular}
\end{table}

Using the fiducial cosmology presented in Table~\ref{tab:cosmologies}, we
compute weights that optimize clustering signal-to-noise for a survey 
with varying density as a function of redshift. Also known as FKP weights
\citep{Feldman94}, we apply a weight to each object,
\begin{equation}
w_{\rm FKP}  =  \frac{1}{1+\bar{n}(z)P_0},
\label{eq:wfkp}
\end{equation}
where $\bar{n}(z)$ is the average comoving density of galaxies as a function 
of redshift and $P_0$ is the value of the power spectrum at scales 
relevant for our study ($k \sim 0.14~h$Mpc$^{-1}$, \citealt{Font14b}). 
For the eBOSS LRG sample we adopt a value 
of $P_0 = 10^4~h^{-3}{\rm Mpc}^3$ which is the same value used in the 
final BOSS CMASS clustering measurements. 
The effective volume is defined as
\begin{equation}
V_{\rm eff} = \int_{z=0.6}^{z=1.0} 
\left(\frac{\bar{n}(z)P_0}{1+\bar{n}(z)P_0} \right)^2 
{A_{\rm eff}} R^2(z) {\rm d}R(z),
\label{eq:veff}
\end{equation}
where $R(z)$ is the comoving distance to redshift $z$ and $A_{\rm eff}$ 
is the effective area (in steradians) of the survey. 

\begin{figure}[t]
\centering
\includegraphics[width=0.45\textwidth]{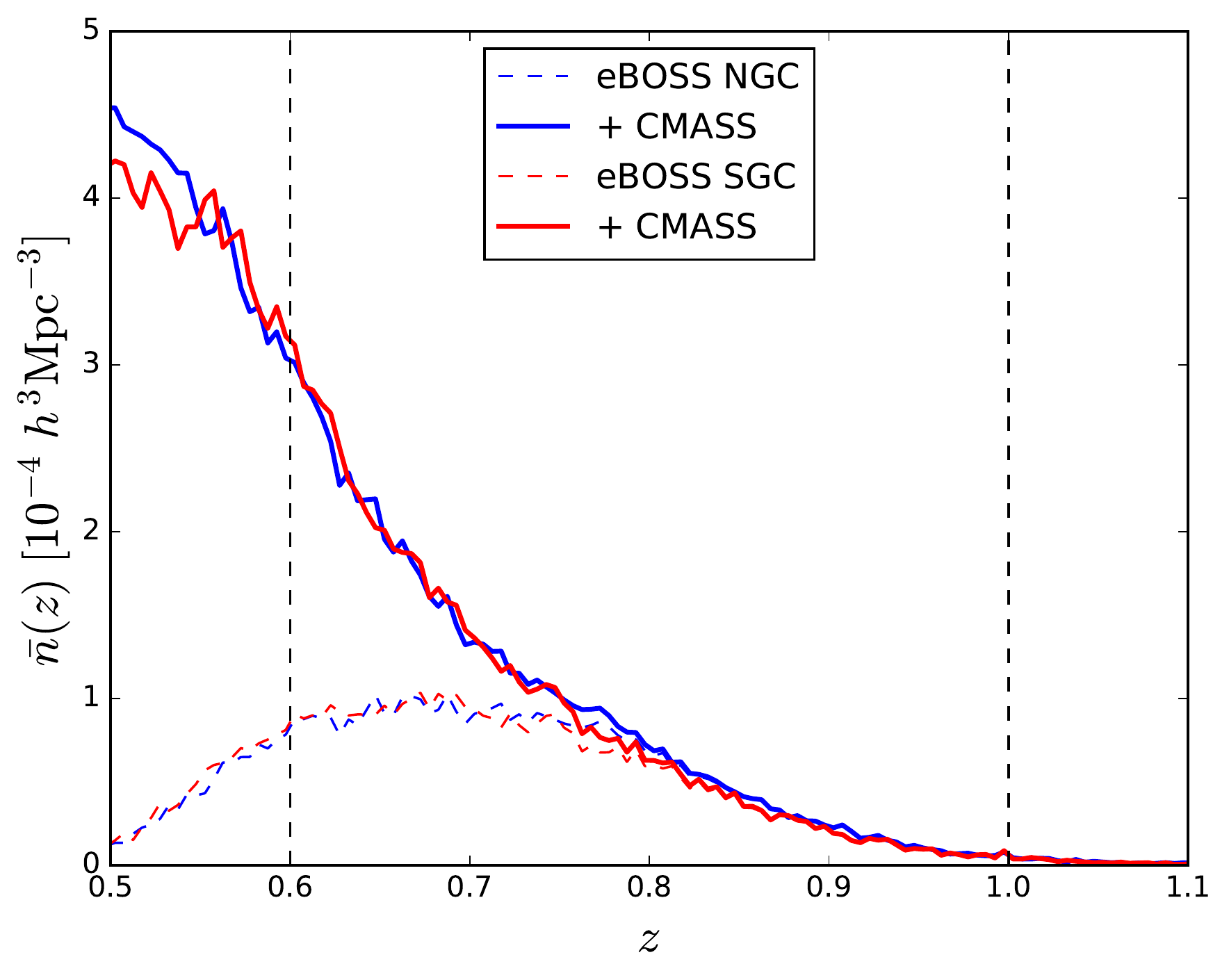}
\caption{Density of LRGs as a function of redshift. 
Dashed vertical lines indicate the redshift range used in our 
clustering measurement. Here we see that a significant fraction of 
eBOSS LRGs have redshifts below 0.6, where CMASS are more numerous. We 
can remark the importance of the CMASS sample between redshift 0.6 and 
0.7.}
\label{fig:nbar}
\end{figure}

We built the random catalog using a sample 50 times that of the galaxy 
sample size.
We applied the same veto masks as for the observed targets. 
Redshifts are assigned to each random point such as to match
the redshift distribution $\bar{n}(z)$ of the
data. Fig.~\ref{fig:nbar} shows the redshift distribution of our
samples, separately for the NGC and the SGC.  
We restrict our analysis to $z>0.6$ to avoid a larger overlap 
with the CMASS sample while not reducing the effective redshift. 
The cut at $z<1.0$ was chosen to avoid low number density of LRGs.  
In section~\ref{sec:systematics}, we describe how systematic effects 
caused by target selection and redshift failures are corrected using 
the same random catalog.

\subsection{Mock catalogs} \label{sec:mocks}

We created a set of mock catalogs, each reproducing the
angular and redshift distribution of galaxies in the DR14 sample
(Fig.~\ref{fig:footprint}), as well as the large-scale correlation
function predicted from the fiducial cosmology. 
We produced simulations for both eBOSS and for the 
combined CMASS+eBOSS sample using redshift distributions shown in 
Fig.~\ref{fig:nbar}.

Mock catalogs were created with the Quick Particle Mesh (QPM) 
method \citep{White14}, also used in recent clustering studies 
\citep{Alam17, Ata17}.
Each realization consists of a different set of 2nd order 
Lagrangian Perturbation Theory (2LPT) initial conditions computed
at $z=25$. These perturbations were evolved to $z=0.7$ 
using a low-force and low-mass
resolution particle-mesh N-body simulation, with time steps of 15\% 
in the log of the scale factor. 
The runs employed here were based on a 2560~\hmpc\ side box containing 
$1280^3$ dark-matter particles. Halos were defined using a 
friends-of-friends algorithm with a linking length of 20\% the mean 
inter-particle spacing. These halos are populated with galaxies 
following a
Halo Occupation Distribution (HOD) model, derived from the small-scale
clustering of the same LRG sample \citep{Zhai17}.  We sub-sample
galaxies in order to reproduce the redshift distribution $n(z)$ and
the angular fiber-completeness as measured from the data. We introduce
redshift failures into our mock catalogs by sampling from the model
derived in section~\ref{sec:redshiftfail}.

\section{Correcting non-cosmological density fluctuations}
\label{sec:systematics}

As described in the previous section, eBOSS LRG targets
were selected to be strictly fainter in $i$-band magnitudes 
than the CMASS sample. Larger photometric errors in this regime 
create a higher rate of contamination by stars and, for some
galaxy spectra, scatter from faint galaxies into the selection. 
Therefore, the eBOSS LRGs are more susceptible to contamination by 
inhomogeneities in target selection and 
by patterns in redshift failures than the CMASS galaxies from BOSS. I
n this section 
we introduce methods to account for this contamination. 
All work in this section is focused on the eBOSS sample
only.

\subsection{Systematics due to photometry}
\label{sec:syst_photo}

In BOSS clustering studies, it was found that galaxy density
is correlated with stellar density and seeing 
\citep{Ho12, Ross11, Ross12, Ross17}. 
These correlations contaminate our clustering
measurements by introducing large-scale power not associated with the 
true distribution of galaxies.  In \cite{Ross17},
systematic weights were computed for each galaxy in order to counteract
these dependencies, but assuming different systematics are
independent.  In this work, we drop the assumption of independent
systematics and use a multiple linear regression, similar to that used
to assess variations in the target selection of LRGs \citep{Prakash16},
quasars \citep{Myers15} and ELGs \citep{Delubac17, Raichoor17}. 
The multiple linear regression has
the advantage of automatically accounting for correlated systematics,
e.g., stellar density and Galactic extinction. In performing 
the regression, we include only galaxies with confident spectroscopic 
redshifts in the region of interest $0.6<z<1.0$. 
The NGC and SGC samples are analyzed independently.

The multiple linear regression calculates a ``density'' model,
$\delta_{\rm phot}$, as a linear combination of maps $m_i$:
\begin{equation}
\delta_{\rm phot}({\rm RA, Dec}) = 
p_0 + \sum_{i=1}^n p_i  m_i({\rm RA, Dec}), 
\end{equation}
where $p_0$ is the average density over the full footprint and
$p_i$ (with $i>0$) are fitted coefficients that minimize 
$\chi^2=(\delta -\delta_{\rm phot})^2/\sigma_{\delta}^2$.
Each map $m_i$ is produced using Healpix with pixels of equal 
area of 189~arcmin$^2$. The observed fluctuations $\delta$ are
estimated from the data (normalized ratio of number of galaxies 
and randoms) as a function of a given systematic value. 
The number of 
galaxies and randoms in each bin are weighted by $w_{\rm FKP}$ 
(Eq.~\ref{eq:wfkp}) in order to account for the cosmological 
fluctuations. Error bars are assumed to be Poisson on the weighted
 number of galaxies per bin. 

\begin{sidewaysfigure}[ht]
\centering
\includegraphics[width=\textwidth]
{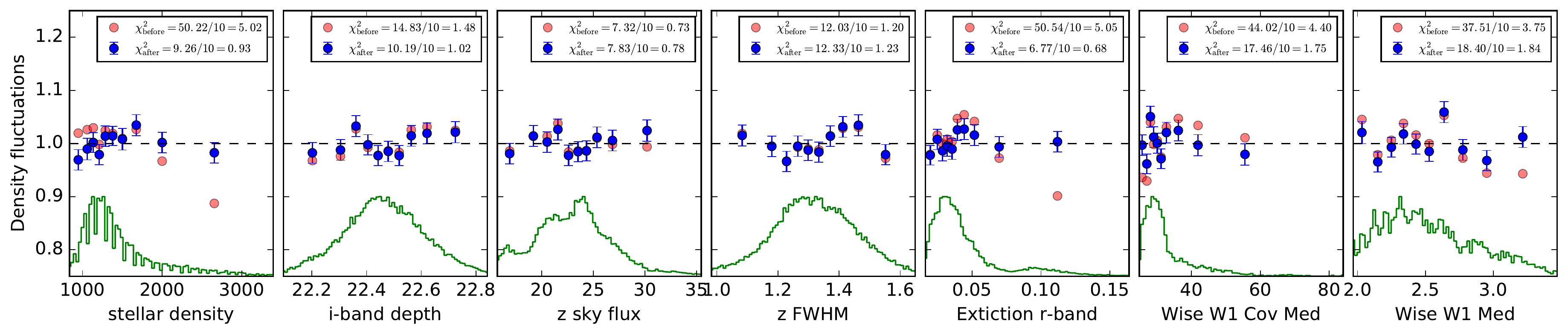}
\includegraphics[width=\textwidth]
{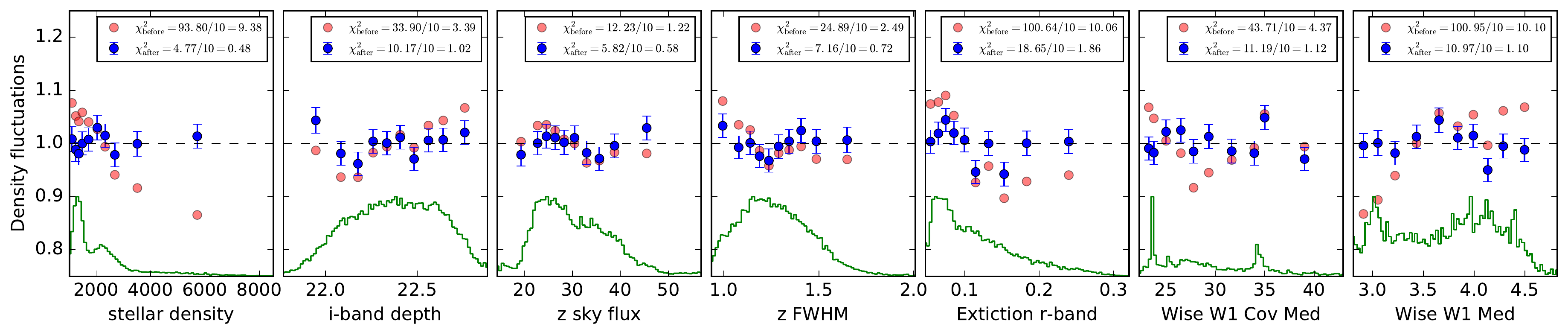}
\caption{Fluctuations of galaxy number density with respect to
  different photometric quantities for the NGC (top panels) and 
  the SGC (bottom panels).  
  Red (blue) points indicate
  fluctuations before (after) corrections. The distribution of the
  same quantities among galaxies is shown in green.  The fit is made
  simultaneously across all maps, accounting automatically for
  correlations between different maps. }
\label{fig:syst_regression}
\end{sidewaysfigure}

Fig.~\ref{fig:syst_regression} shows the result of the regression
using seven different maps. Five of these maps are derived from SDSS 
photometry \citep{Doi10, DR8}: stellar density, $i$-band depth, 
$z$-band sky flux, $z$-FWHM and $r$-band extinction, while two maps
are derived from WISE photometry \citep{Wright10, Lang14}: median number of single-exposure
frames per pixel in WISE W1 band (WISE W1 Cov Med) and median of
accumulated flux per pixel in the WISE W1 band (WISE W1 Med). 
Since different SDSS bands  \citep{Fukugita96} are strongly correlated, we restrict 
our analysis to a single band per systematic. 
In the NGC, $\chi^2/dof = 192.2/(70-8)$ before corrections and 
$80.6/(70-8)$ after corrections, while 
in the SGC we obtain $\chi^2/dof = 365.8/(70-8)$ before corrections 
and $68.2/(70-8)$ after corrections. 
The most important improvements are related
to dependencies with stellar density, extinction and WISE quantities.
This analysis can be perfomed in many redshift bins, 
however, no further improvement was obtained. Hereafter,
we compute systematic maps using a single redshift bin.

Once the density model is derived, our sample can, in principle, be
corrected either by applying to the galaxies a set of systematic
weights defined as $w_{\rm sys} = 1/\delta_{\rm phot}$, or by
sub-sampling the random catalog  as a function of RA and Dec 
to mimic the density model.

\begin{figure}[t]
\centering
\includegraphics[width=0.45\textwidth]
{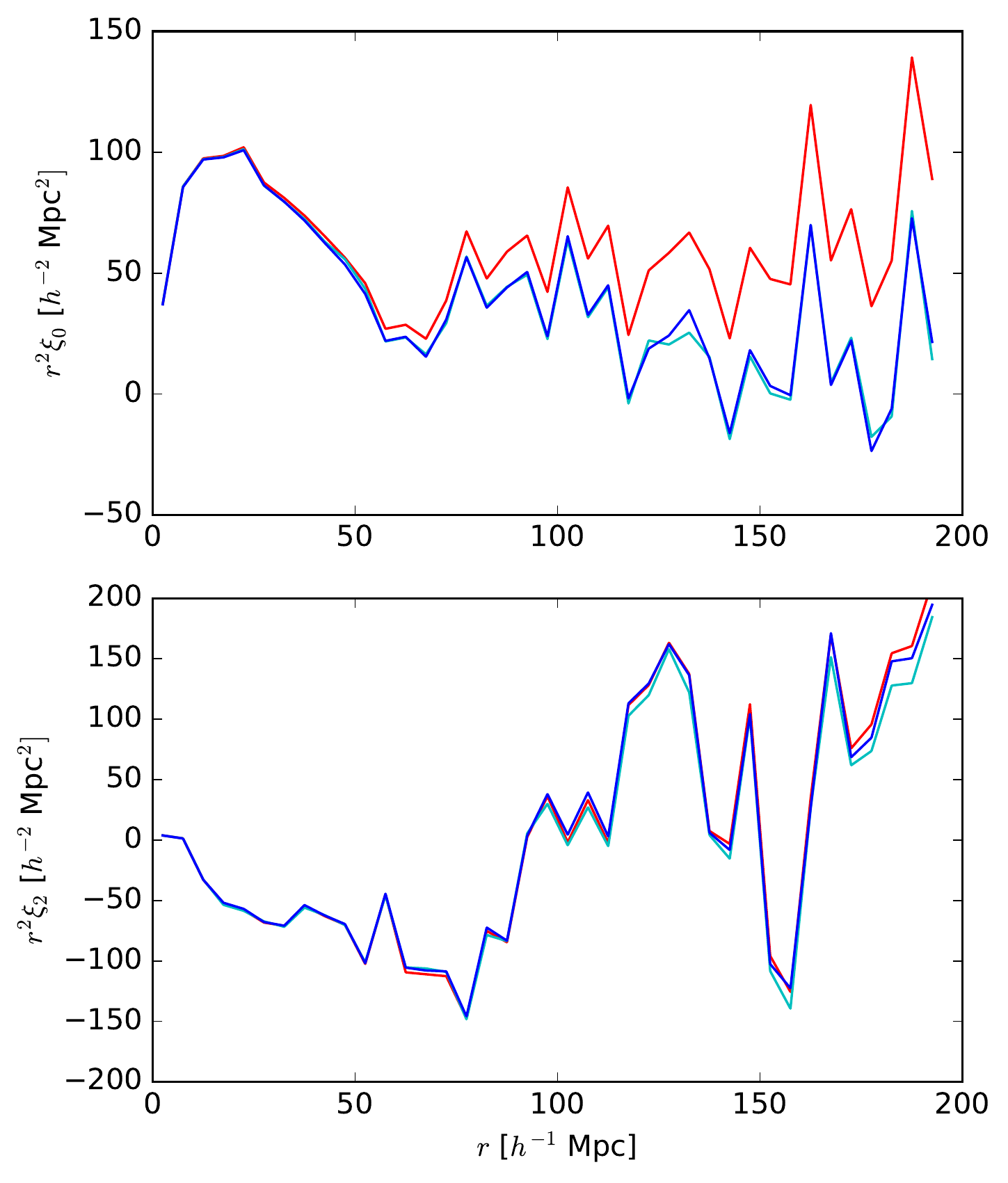}
\includegraphics[width=0.45\textwidth]
{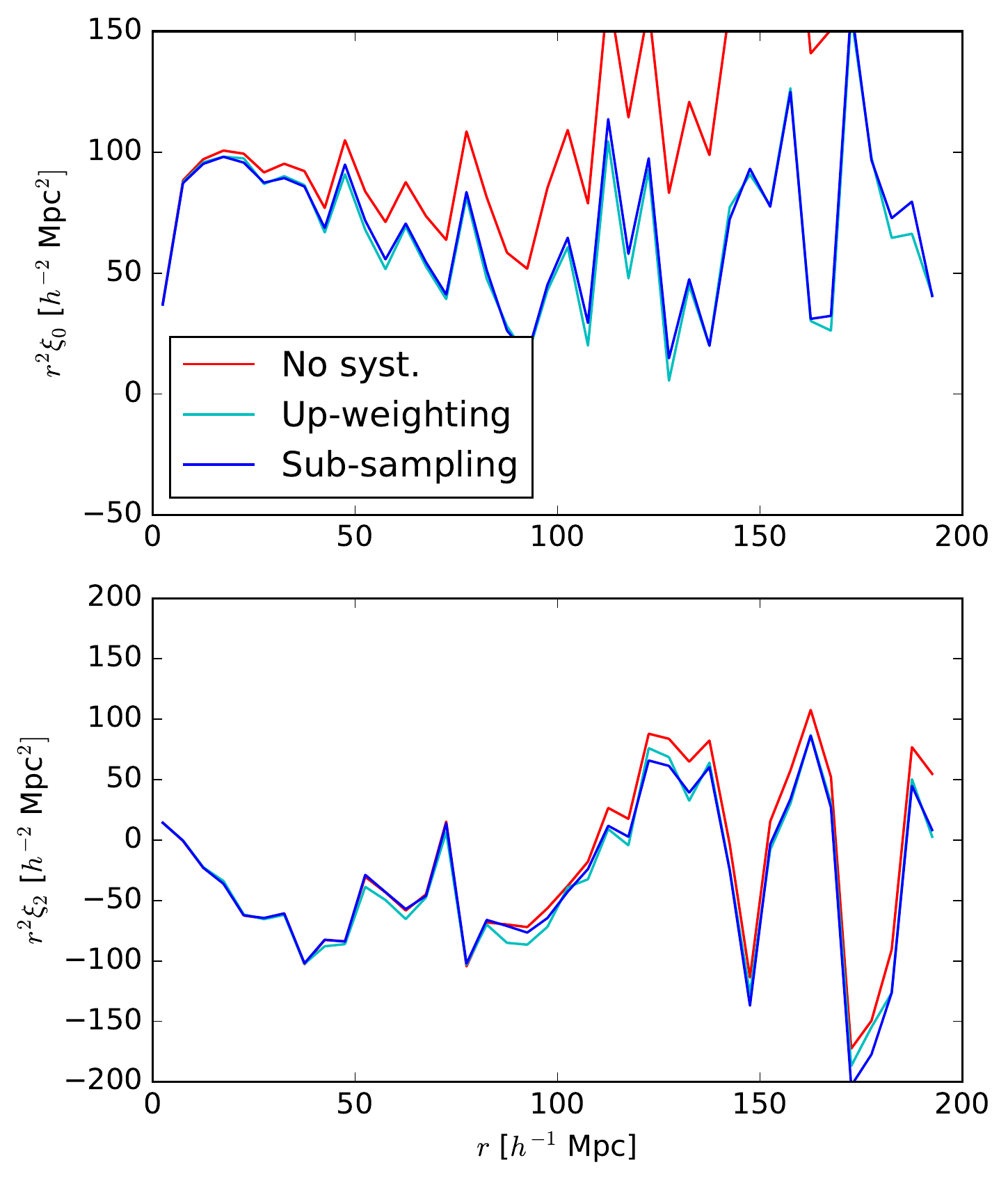}
\caption{Monopole (top panels) and quadrupole (bottom) of 
the correlation function for NGC (left) and SGC (right) 
before (red lines) and after correcting 
for targeting systematics. The cyan line shows
the result when using the up-weighting scheme and 
the blue line shows the sub-sampling of randoms.
Since our corrections for redshift failures are also 
applied on the random catalog, we employ hereafter 
the sub-sampling of randoms as our fiducial 
method to correct for targeting systematics.}
\label{fig:syst_methods}
\end{figure}

Fig.~\ref{fig:syst_methods} presents the monopole and the quadrupole of
the correlation function calculated using the standard \citet{LS93}
estimator. 
We show the observed correlation function before accounting for
non-cosmological fluctuations in target density and compare the 
results of correcting with an up-weighting scheme to results
from sub-sampling of randoms. 
Correlations are biased positive even for large separations 
where no cosmological signal is expected. As we can see in the figure, 
the quadrupole is barely affected by 
this kind of systematic error.

Differences between the two correction techniques are smaller than
error bars and are consistent with being caused by the relatively
smaller number of randoms for the sub-sampling case. 
Hereafter, targeting systematics
are corrected by sub-sampling of randoms.

\subsection{Correcting for redshift failures}
\label{sec:redshiftfail}

\begin{figure}
\centering
\includegraphics[width=0.4\textwidth]{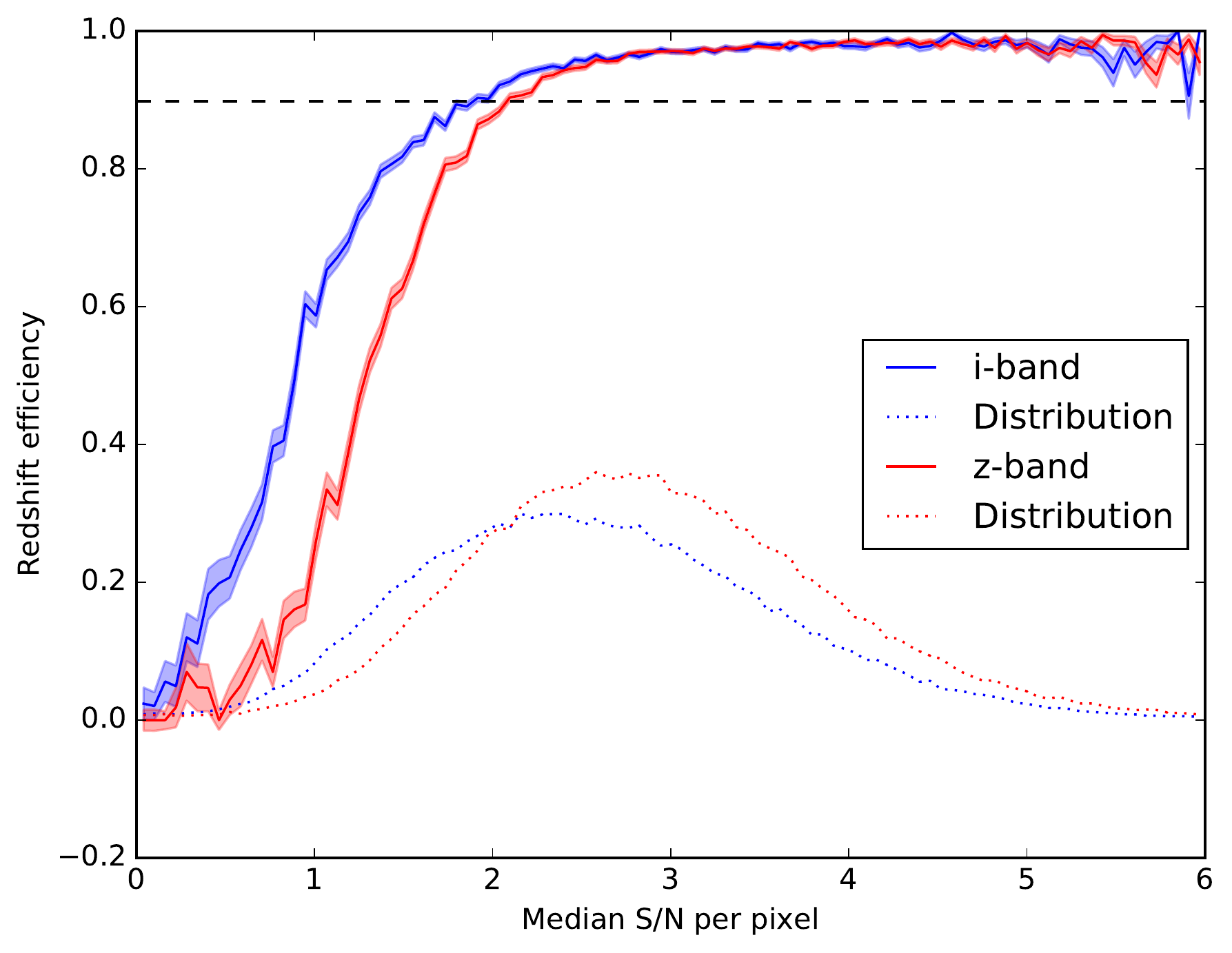}
\caption{Average redshift efficiency as a function of median pixel
  signal-to-noise ratio in the $i$ (blue line) and $z$ 
  (red line) SDSS bands.
  Dotted lines indicate the distribution of median S/N of the 
  eBOSS LRG sample. 
  The dashed line represents the average redshift efficiency of the whole
  sample of 90.5\%.}
\label{fig:eff_vs_snmedian}
\end{figure}

Variations in the quality of spectroscopy can have a similar effect on
clustering as variations in the quality of the photometry used to identify
targets.   
Fig.~\ref{fig:eff_vs_snmedian} reveals that lower S/N spectra yield, on average,
fewer statistically confident redshifts. We define redshift efficiency as
\begin{equation}
\eta = \frac{N_{\rm gal}}{N_{\rm zfail}  + N_{\rm gal}},  
\end{equation}
where quantities are defined in Section~\ref{sec:catalog}. 
If the expected distribution
of failures is not uniformly distributed across the sky, our
clustering measurements might be biased.  In this section, we 
introduce 
a new technique to account for these failures.  Using mock catalogs
(described in Section~\ref{sec:mocks}), we demonstrate that our method 
leads to unbiased clustering measurements.

In previous studies from BOSS, redshift failures were accounted for 
by an up-weighting technique. These spectra lacking a confident redshift
transfer their weight to the nearest object
in the sky with a confident redshift (either a galaxy or a star). 
In other words, the nearest neighbor of a failure will count double
when counting pairs.
This simple correction procedure leads to unbiased results on scales far 
larger that the average separation between a failure and a non-failure and 
as long as the rate of redshift failures is low.  
For example, in the BOSS CMASS sample the overall failure rate was 1.8\%
and the median separation was 3.8\,arcmin. In the eBOSS LRG sample,
the failure rate is around 10\% and the median separation is
5\,arcmin. The higher failure rate and larger average separation 
force us to revisit the manner in which redshift incompleteness is 
addressed in eBOSS.

Instead of using the up-weighting technique, we 
derive a model describing the probability of an observation
of a target galaxy yielding a confident redshift. In our model, this
probability depends on the the position of its fiber in the focal
plane and the overall signal-to-noise ratio of the plate.
The model for failures is then applied to the random sample, by
sub-sampling, mimicking the patterns retrieved in our model.

\begin{figure}
\centering
\includegraphics[width=0.45\textwidth]{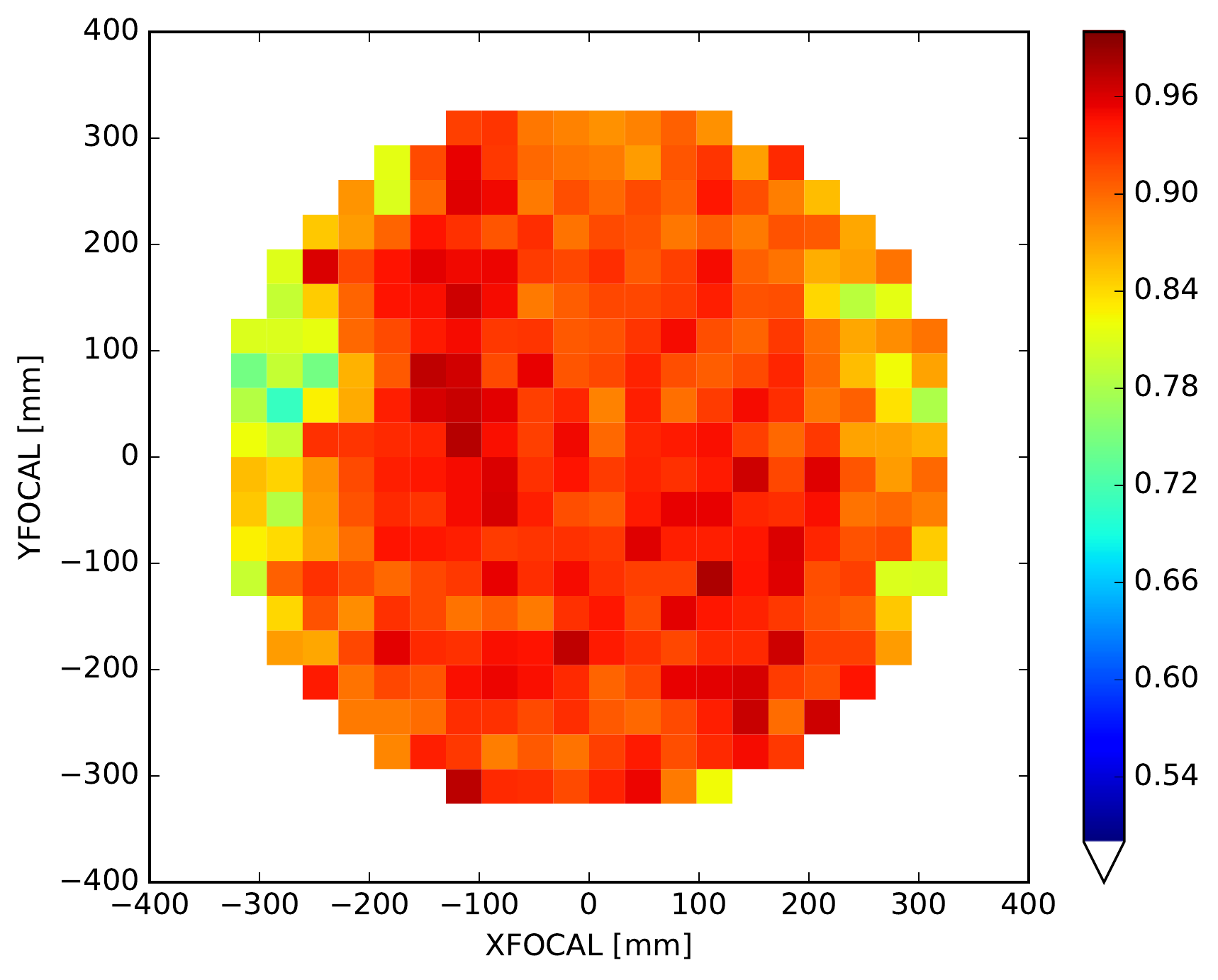}
\includegraphics[width=0.45\textwidth]{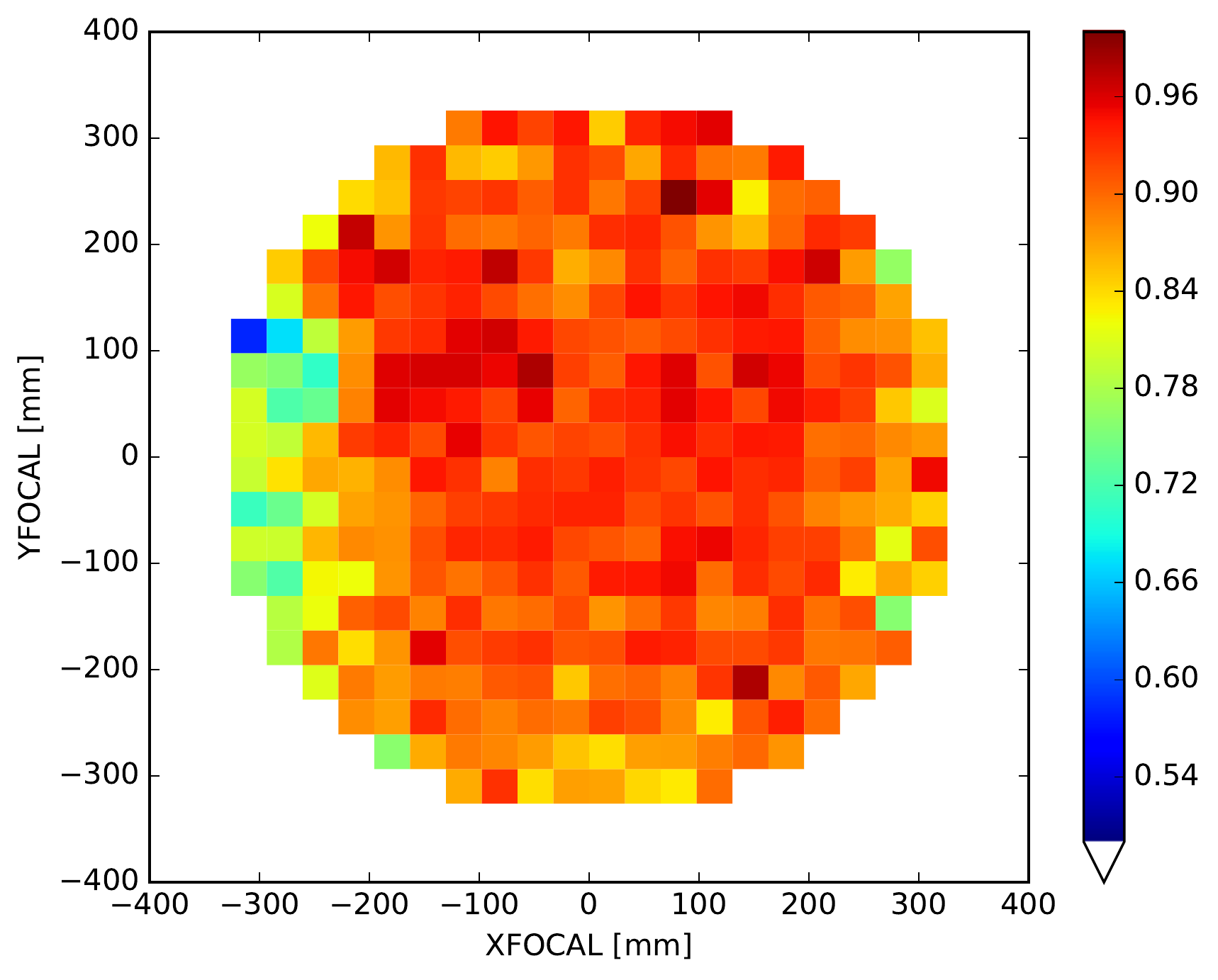}
\caption{Average redshift efficiency as a function of physical 
position of the optical fiber in the focal plane. Left (resp. right) 
panel shows the NGC (resp. SGC) failure rates. The region where 
YFOCAL is negative correspond to spectrograph \#1 while YFOCAL $>0$ 
corresponds to spectrograph \#2.}
\label{fig:eff2d_xy}
\end{figure}

Fig.~\ref{fig:eff2d_xy} shows the probability of obtaining a confident
redshift (hereafter called the redshift efficiency) for a galaxy as a
function of its position in the focal plane. We observe a decrease in
this probability near the side edges of the focal plane.  The reason
for this behavior is that the light transmitted through fibers near the 
side-edges of the focal plane arrive near the edges of the CCD, where the 
optical performance inside the spectrographs is degraded, leading to a 
larger point spread function and optical aberrations such as coma 
\citep{Smee13}.  
By associating each random with a plate and location within the plate, 
we can use the binned probabilities to
sub-sample the random catalog. We divide the focal plane into 20 bins 
across the diameter of the focal plane in cartesian coordinates, 
XFOCAL and YFOCAL, that range from -326 to 326mm.
Our choice of bin size is large
enough to minimize Poisson noise in galaxy counts but small enough to
identify  the failure-rate pattern on the scales of interest.

\begin{figure}[t]
\centering
\includegraphics[width=0.45\textwidth]{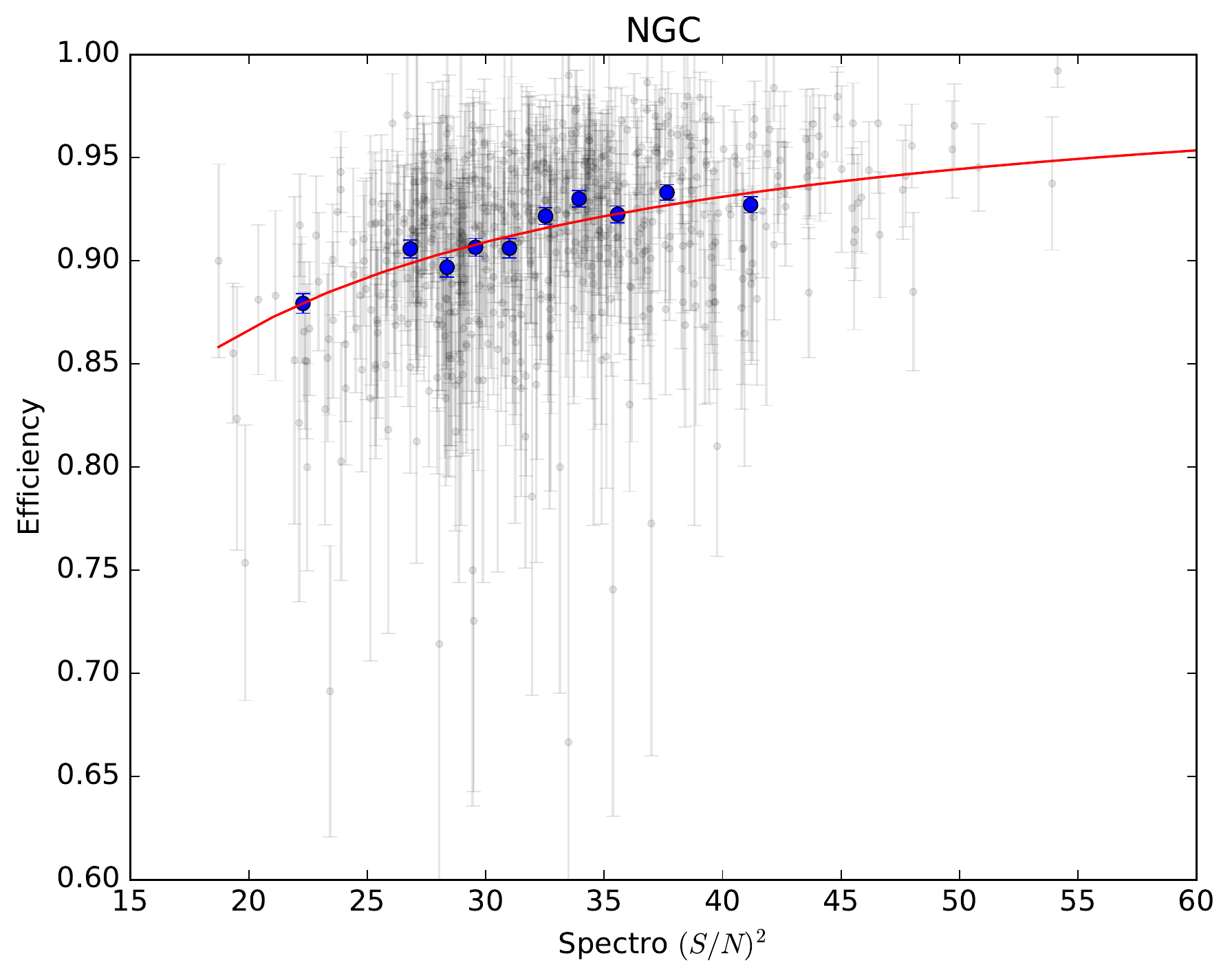}
\includegraphics[width=0.45\textwidth]{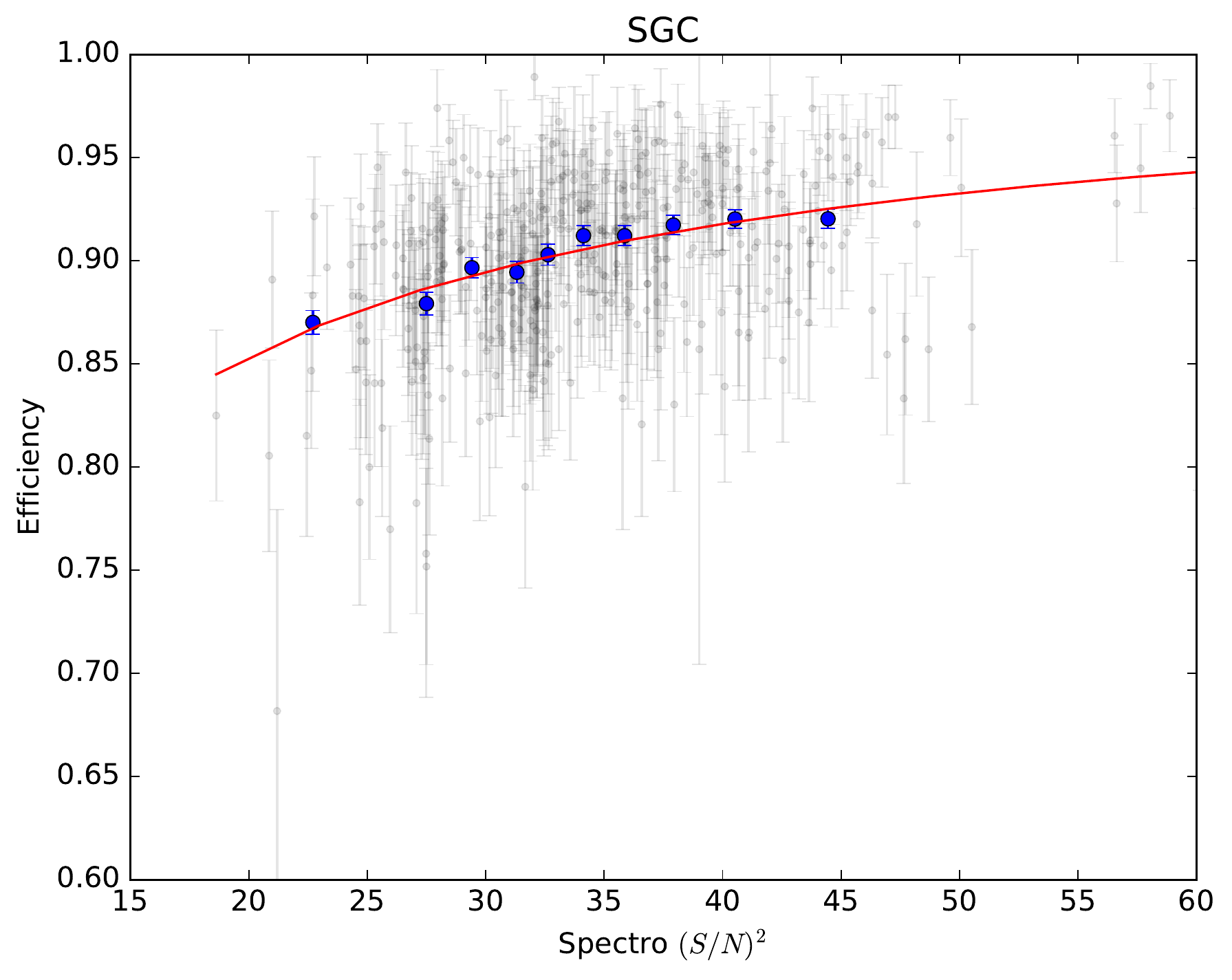}
\caption{Average redshift efficiency per spectrograph as a function of 
its signal-to-noise ratio squared binned (blue points) and unbinned (gray points). 
Red line show the best-fit 
models. The left panel shows the results for the NGC while the right
panel shows the SGC. Signal-to-noise is computed as the median per pixel 
over the i-band wavelength range. }
\label{fig:eff_vs_snplate}
\end{figure}

Fig.~\ref{fig:eff_vs_snplate} presents this dependency of the
redshift efficiency with signal-to-noise ratio of both spectrographs
for each plate.  We use these
half-plates since two independent spectrographs have different 
throughput \citep{Smee13}.  On average, fibers lying on
the YFOCAL$<0$ region (in Fig.~\ref{fig:eff2d_xy}) encounter 
spectrograph \#1, while the others are typically imaged through 
spectrograph \#2.  
Independent measurements of $S/N$ are made per spectrograph and 
per optical band (corresponding to SDSS $g$, $r$, $i$). 
We used $i$ band values only, given that most of the signal of 
LRG spectra is observed in the $i$ band. 
The binned data (blue points) 
are included in order to reduce Poisson noise. 
We fit the efficiency $\eta$ using a simple model,
\begin{equation}
\eta(x) = \frac{1}{1+1/p(x)},
\label{eq:eff_vs_platesn}
\end{equation}
where $x=(S/N)^2$ and $p(x)$ is a first order polynomial. This
choice of model ensures the correct asymptotic behavior when 
$S/N\rightarrow 0$ or $S/N \rightarrow \infty$. 
The best-fit models yield $\chi^2/{\rm dof} = 47.1/(20-2)$ for the NGC,
and $\chi^2/{\rm dof} = 39.2/(20-2)$ for the SGC. 
These values are higher than unity, indicating that efficiencies 
may depend on factors other than spectrograph $S/N$. 
However, we were not able to identify any other significant factors. 
We expect to improve the model of redshift failures with a larger data
sample. 

Our final efficiency model is the product of the two efficiencies
given in Fig.~\ref{fig:eff2d_xy} and Eq.~\ref{eq:eff_vs_platesn}. We
normalize the efficiencies such that the final product is consistent
with the average spectrograph efficiency given by the red line 
in Fig~\ref{fig:eff_vs_snplate} (since the latter already includes 
the average focal plane efficiencies).

The sub-sampling technique is implemented as follows.
For each random galaxy, we assign a plate, XFOCAL and YFOCAL values 
based on its location in the sky. In overlap regions covered by more 
than a single plate, a random plate value among those observing this
region is assigned to this random galaxy. Given the plate number and 
the (RA, Dec) of each random, we can assign a position in the 
focal plane of that plate. We draw random numbers and we remove 
random galaxies based on the probability given by the model. 

In order to test our procedure, we included redshift failures
on the set of 1000 mock catalogs for eBOSS galaxies,
following the model derived from real data. 
We first assign mock galaxies to plates (and their XFOCAL and YFOCAL)
as done with the randoms. For each galaxy, we randomly convert it
as a failure based on the probability given by the model. 
We apply two 
methods to account for these failures: the nearest-neighbor
up-weighting and the sub-sampling of randoms. For the sub-sampling
case, we use redshift failures from the mock itself to derive 
individual redshift efficiency models, employing the same algorithm 
that is used to derive the inefficiency model from the data. 
Doing so accounts for the Poisson noise
that could be caused by the binned data. 

\begin{figure}
\centering
\includegraphics[width=0.49\textwidth]{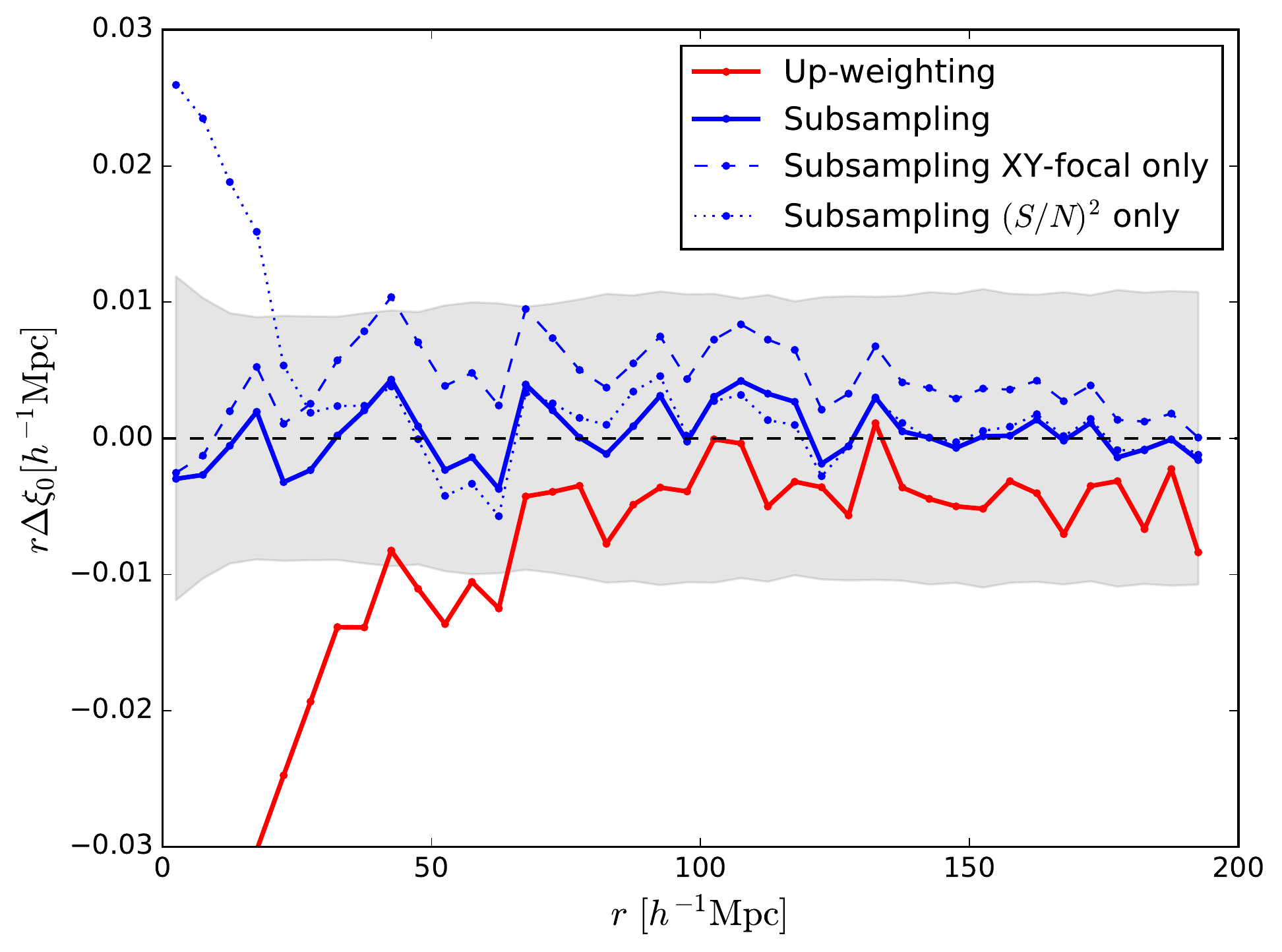}
\includegraphics[width=0.49\textwidth]{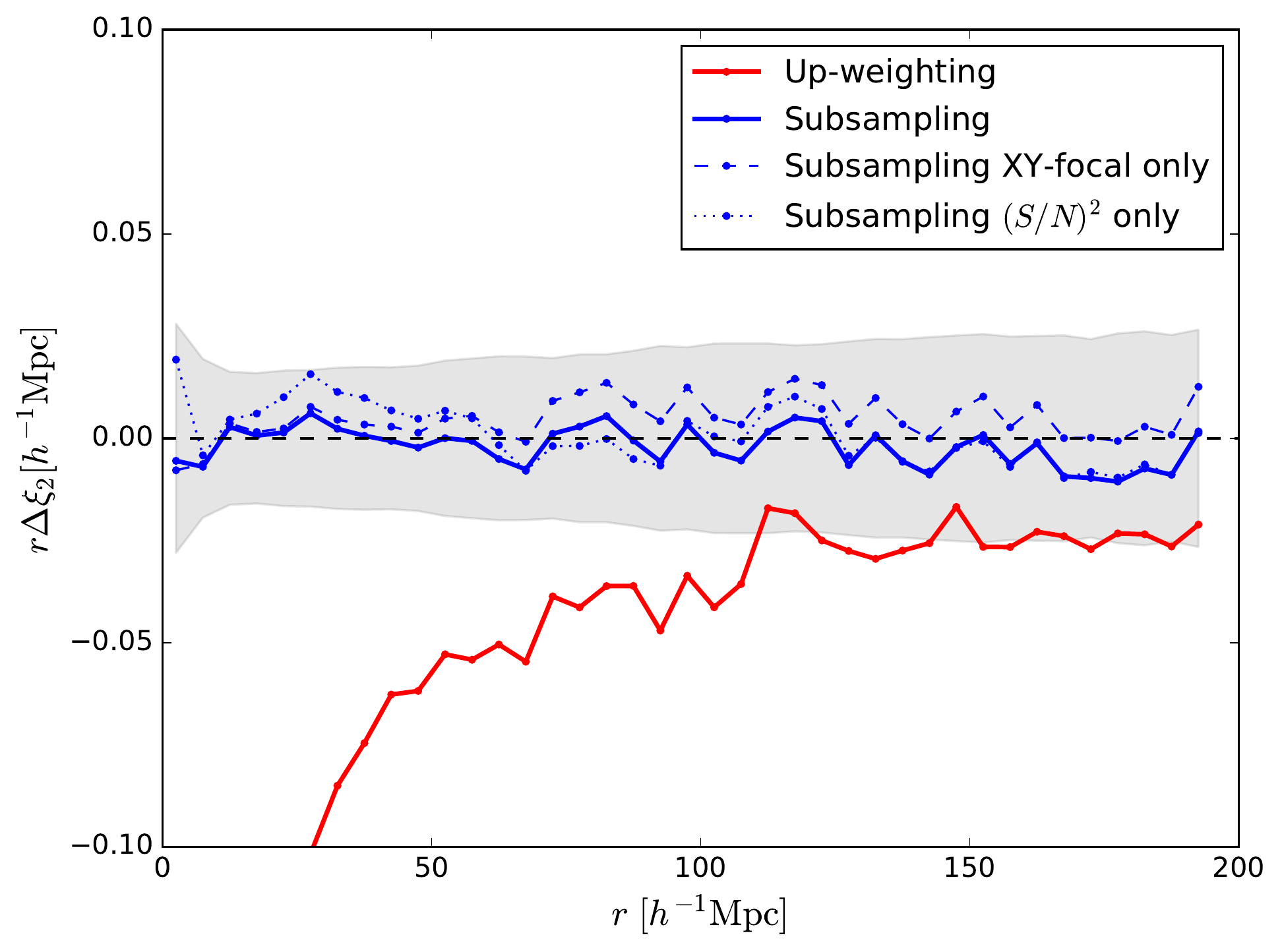}
\caption{Difference between the average mock correlation function where 
redshift failures are corrected and the mocks without synthetic failures. 
The left panel presents the monopole and the right panel shows the 
quadrupole, both scaled by $r$ for clarity. Different lines present the 
results with different schemes. 
The grey band is the error on the mean of 1000 mock catalogs of the 
eBOSS NGC sample (without CMASS). The up-weighting of nearest neighbor 
technique introduces a bias of 5\% in the quadrupole on all scales while
the sub-sampling is unbiased.}
\label{fig:effect_failure_corrections}
\end{figure}

Fig~\ref{fig:effect_failure_corrections} shows the impact of different
correction methods on the average correlation function of the 1000 mock
catalogs.  The reference correlation function is computed from the same
mock catalogs without any synthetic redshift failures. 
The grey region represents the error on the mean  
of 1000 correlation functions (all curves have
similar errors). The nearest-neighbor up-weighting scheme (blue curve) 
introduces structure into the monopole with amplitude smaller than 
the error on the 1000 mocks, except at scales 
below 20\hmpc\ which are usually discarded in BAO analyses. 
However, this scheme introduces a bias of at least 5\% on all
scales in the amplitude of the quadrupole. 
The sub-sampling technique (green lines) 
has better performance than up-weighting for all scales, 
for both monopole and quadrupole, yielding no significant bias at this 
level of precision. 
We also applied two ``non-complete'' versions of the sub-sampling scheme,
where we assume the efficiency model depends only on one factor:
either focal plane position or only spectrograph $S/N$.
Even when our assumed model for 
redshift failures is not complete, our model is superior 
to the up-weighting scheme. 

The bias on the quadrupole introduced by up-weighting would yield 
biased estimates of the growth rate of structures in studies 
of redshift-space distortions. We recommend that future work on RSD using 
galaxy samples containing significant failure rates 
 implement our correction scheme. 

We use the sub-sampling techniques on the eBOSS LRG sample in all 
clustering measurements in this work. For the CMASS $z>0.6$ sample
we employ the weights used in the final BOSS measurements, which use
the up-weighting technique. We combine CMASS and eBOSS samples, 
both data and randoms, making sure that the ratio of number of galaxies 
over randoms is the same. 

\section{The model and fitting methodology}
\label{sec:methods}

In this section, we describe the model used to fit the correlation
function, the reconstruction procedure applied, and tests on
the mock catalogs.

\subsection{The model}
\label{sec:model}

To fit the measured correlation function we follow the
standard approach described in previous papers (e.g., \citealt{Alam17,
  Ata17}).  The model redshift-space correlation function is obtained
from a Fourier transform of the power-spectrum, which is defined as
\begin{equation}
P(k, \mu)  = \frac{b^2 \left[1+\beta(1-e^{-k^2\Sigma_r^2/2})\mu^2\right]^2}
{(1+ k^2\mu^2\Sigma_s^2)}  
\left[ P_{\rm lin}(k) + P_{\rm peak}(k)
\left(e^{-k^2\Sigma_{\rm nl}^2(\mu)}-1\right)   \right]
\label{eq:pk2d}
\end{equation}
where $b$ is the linear bias, $\beta$ is the redshift-space
distortions parameter, $k$ is the modulus of the wave-vector and $\mu$ is
the cosine of the angle between the wave-vector and the line-of-sight.
We introduce anisotropic non-linear broadening of the BAO peak by
multiplying the ``peak-only'' power spectrum $P_{\rm peak}$ by a
Gaussian term with
$\Sigma_{\rm nl}^2(\mu) = \Sigma_\parallel^2 \mu^2 +
\Sigma^2_\perp(1-\mu^2)$. 
The non-linear random motions on small scales are modeled by
a Lorentzian term parametrized by $\Sigma_s$.  Typical values for the
damping terms can be computed by fitting the average of many mocks:
$(\Sigma_\parallel, \Sigma_\perp, \Sigma_s) = (9.5, 6.0, 2.0)$~\hmpc\
for pre-reconstruction and
$(\Sigma_\parallel, \Sigma_\perp, \Sigma_s) = (5.5, 5.5, 0)$~\hmpc\
for post-reconstruction. Given the relatively low signal-to-noise ratio
of the correlation functions of this sample, all fits have 
$(\Sigma_\parallel, \Sigma_\perp, \Sigma_s)$ values fixed.
Following theoretical motivation of \citet{White15} and \citet{Seo16}, 
we apply a term $S(k) = e^{-k^2\Sigma_r^2/2}$ to the post-reconstruction
modeling of the correlation function. This term models the smoothing used 
in our reconstruction technique, where $\Sigma_r = 15$\hmpc\ 
(see section~\ref{sec:recon}). 
This term was used in the BOSS DR12 results of \citet{Ross17} and
 \citet{Beutler17a}. 
We follow \cite{Kirkby13} to compute
$P_{\rm peak}$ from the linear power-spectrum $P_{\rm lin}$, by
computing its correlation function, fitting a third-order polynomial
function over the peak region, and transforming back to Fourier space.
The linear power spectrum $P_{\rm lin}$ is computed using the code 
CAMB\footnote{\url{camb.info}} \citep{Lewis2000} with cosmological 
parameters of 
our fiducial cosmology (Table~\ref{tab:cosmologies}).

In practice, we derive multipoles for the correlation function 
$\xi_\ell(r)$ from multipoles of the power-spectrum $P_\ell(k)$, 
simply defined as:
\begin{equation}
P_\ell(k) = \frac{2\ell+1}{2} \int_{-1}^{1} P(k, \mu) L_\ell(\mu) ~{\rm d}\mu
\end{equation}
where $L_\ell$ are Legendre polynomials. The correlation function is:
\begin{equation}
\xi_\ell(r) = \frac{i^\ell}{2\pi^2}\int_0^\infty 
k^2 j_\ell(kr) P_\ell(k) ~ {\rm d}k
\end{equation}
where $j_\ell$ are the spherical Bessel functions.

In order to measure the BAO peak position, we scale separations
$r$ with an isotropic dilation factor, $\aiso$, defined as the
ratio of the ``spherically averaged'' distance:
\begin{equation}
D_V(z) = [zD_M^2(z)D_H(z)]^{1/3}
\end{equation}
to the sound horizon scale $r_d$, normalized to its value 
in our fiducial cosmology, i.e:  
\begin{equation}
\aiso = \frac{D_V(z_{\rm eff})/r_d}
{D^{\rm fid}_V(z_{\rm eff})/ r_d^{\rm fid}}.
\label{eq:alpha_iso}
\end{equation}

Another choice of parametrization of the BAO peak position uses 
two dilation parameters that decompose the scaling into
transverse, $\aperp$, and radial, $\apar$, directions. 
These quantities are related,
respectively, to the comoving angular diameter distance, 
$D_M = (1+z)D_A(z)$, and to the Hubble distance, $D_H = c/H(z)$, by
\begin{equation}
\aperp = \frac{D_M(z_{\rm eff})/r_d}
              {D_M^{\rm fid}(z_{\rm eff})/ r_d^{\rm fid}}
 \hspace{2cm} 
\apar =  \frac{D_H(z_{\rm eff})/r_d}
              {D_H^{\rm fid}(z_{\rm eff})/ r_d^{\rm fid}}
\label{eq:aperp_apar}
\end{equation}
In our implementation, we apply the scaling factors in the 
two-dimensional power spectrum (Eq.~\ref{eq:pk2d}) before 
computing its multipoles and the associated Hankel transforms (e.g.
\citealt{Beutler14, GilMarin17}).

Unknown systematic effects in the survey might create large-scale
correlations that could contaminate our measurements. We take into account
any spurious correlations by introducing into our model a smooth
functions of separation. Our final template can be written as
\begin{equation}
\xi^t_\ell(r) = \xi_\ell(\alpha r) + a_{\ell 0} + \frac{a_{\ell 1}}{r}
     + \frac{a_{\ell 2}}{r^2}.
\label{eq:template}
\end{equation}

For isotropic fits, we only fit the monopole of the correlation function,
fixing the value of $\beta = 0.3$.
For anisotropic fits, we fit both monopole and quadrupole, 
leaving $\beta$ as a free parameter and fitting $b$ with a Gaussian prior
of 30\% around $b=2.3$. For all fits, the broadband parameters are
free while dilation parameters are fitted in the range $0.5 < \alpha < 1.5$.
A total of 5 parameters are fitted on isotropic fits and 
10 parameters for anisotropic fits.

\subsection{Parameter estimation}
\label{sec:parameter_est}

The best-model is found by minimizing $\chi^2 = d \hat{W} d^T$, 
where $d$ is the monopole (and quadrupole) of the correlation 
function, and $\hat{W}$ is the estimated
precision matrix, defined as the inverse of the estimated 
covariance matrix $\hat{C}$. We use the iMinuit python 
package\footnote{\url{http://iminuit.readthedocs.io}} that implements
a quasi-Newton method using DFP formula to find minima.
All of our covariance matrices are derived from the scatter of 
measurements from mock catalogs. 
In order to account for the finite number of mock measurements 
used to derive
$\hat{C}$, the unbiased estimator for the precision matrix 
\citep{Hartlap07, Taylor14,Percival2014} should be written as
\begin{equation}
\hat{W} = \left(1 - \frac{n}{N_{\rm mocks}-1}\right)\hat{C}^{-1},
\label{eq:precision_matrix}
\end{equation}
where $n$ is the number of elements in the data vector $d$.

Errors from best-fit parameters are derived from the 
$\Delta \chi^2 =1$ region of the marginalized $\chi^2$
profiles. Errors on the BAO scale $\aiso$ are usually asymmetric 
for the current signal-to-noise ratio of our samples.

\subsection{Reconstruction}
\label{sec:recon}

To reduce the non-linear effects on the BAO feature, we
applied ``reconstruction'' \citep{Eis07, Burden15, Magana17} to the
eBOSS+CMASS sample.  The reconstruction method reverses a fraction of
non-linear motion of the overdensities, sharpening the BAO peak and
thus increasing the precision of our distance-redshift measurement.
 
Using the Zeldovich approximation, we calculate Lagrangian
displacements $\vec{\Psi}$ based on an estimate of the velocities made 
from the density field.  In order to estimate the galaxy density field, 
we insert the NGC or the SGC region inside a box with width 3.6 comoving
$h^{-1}$Gpc, assigning the galaxies and randoms into a grid of
$512^3$ cells (using our fiducial cosmology). 
We use the cloud-in-cell scheme to
assign galaxies and randoms to the grid\footnote{A discussion 
on the effects of assignment scheme on Fast Fourier Transforms 
can be found in \citet{Cui08}. }. The density is smoothed using
a three-dimensional Gaussian kernel with 15\hmpc\ as its
characteristic length. \citet{Burden14, Magana17} studied how results vary as a
function of this smoothing length and found that 15\hmpc\ is close to
optimal in terms of sharpening the BAO peak for densities similar to
those matching our samples.  

For a redshift-space density field, we
find an approximate solution for the displacement field using 
Inverse Fast Fourier Transforms (IFFT), 
 \begin{equation}  
 \vec{\Psi} = {\rm IFFT}\left[ \frac{-i \vec{k} 
 \tilde{\delta}(\vec{k})}{bk^2}\right]  
 - \frac{f}{b+f} \left( {\rm IFFT}
  \left[ \frac{-i \vec{k} \tilde{\delta}(\vec{k})}{bk^2} \right] 
  \cdot \hat{r} \right) \hat{r},
  \label{eq:psi_fft}
 \end{equation}
 where $\tilde{\delta}$ is the Fourier transform of the density field,
 $\vec{k}$ is the wave-vector, $b$ and $f$ are values for the bias and
 growth rate for the sample, and $\hat{r}$ is the unit vector in the
 radial direction. Because the overdensity field in redshift-space is
 not irrotational, this equation is only an approximation, 
 although it can be
 made more accurate, tending towards the true solution, using an
 iterative approach \citep{Burden15}.

 Once the displacement field has been calculated, we move both data
 and randoms by the expected value to obtain the reconstructed
 catalogs. We can also remove redshift-space distortions by moving
 galaxies by an additional displacement defined as
\begin{equation}
\vec{\Psi}_{\rm RSD} = -f ( \vec{\Psi}\cdot \hat{r}) \hat{r}.
\end{equation}
This action does not reduce the signal-to-noise ratio of the
reconstructed overdensity field measurement (the transition from real
to redshift-space increases the information content). 

We consider three implementations of the reconstruction method. 
The differences between them are whether or not we remove RSD, and how
many iterations we perform as in \citet{Burden15}. 
\begin{itemize}
\item[] A: Not removing RSD and not performing iterations
\item[] B: Removing RSD and not performing iterations
\item[] C: Removing RSD and performing iterations
\end{itemize}

For case B, we use Eq. 19 in \cite{Burden15}. 

\subsection{Fitting mock catalogs}
\label{sec:mock_fits}

\begin{figure}[t]
\centering
\includegraphics[width=0.49\textwidth]
{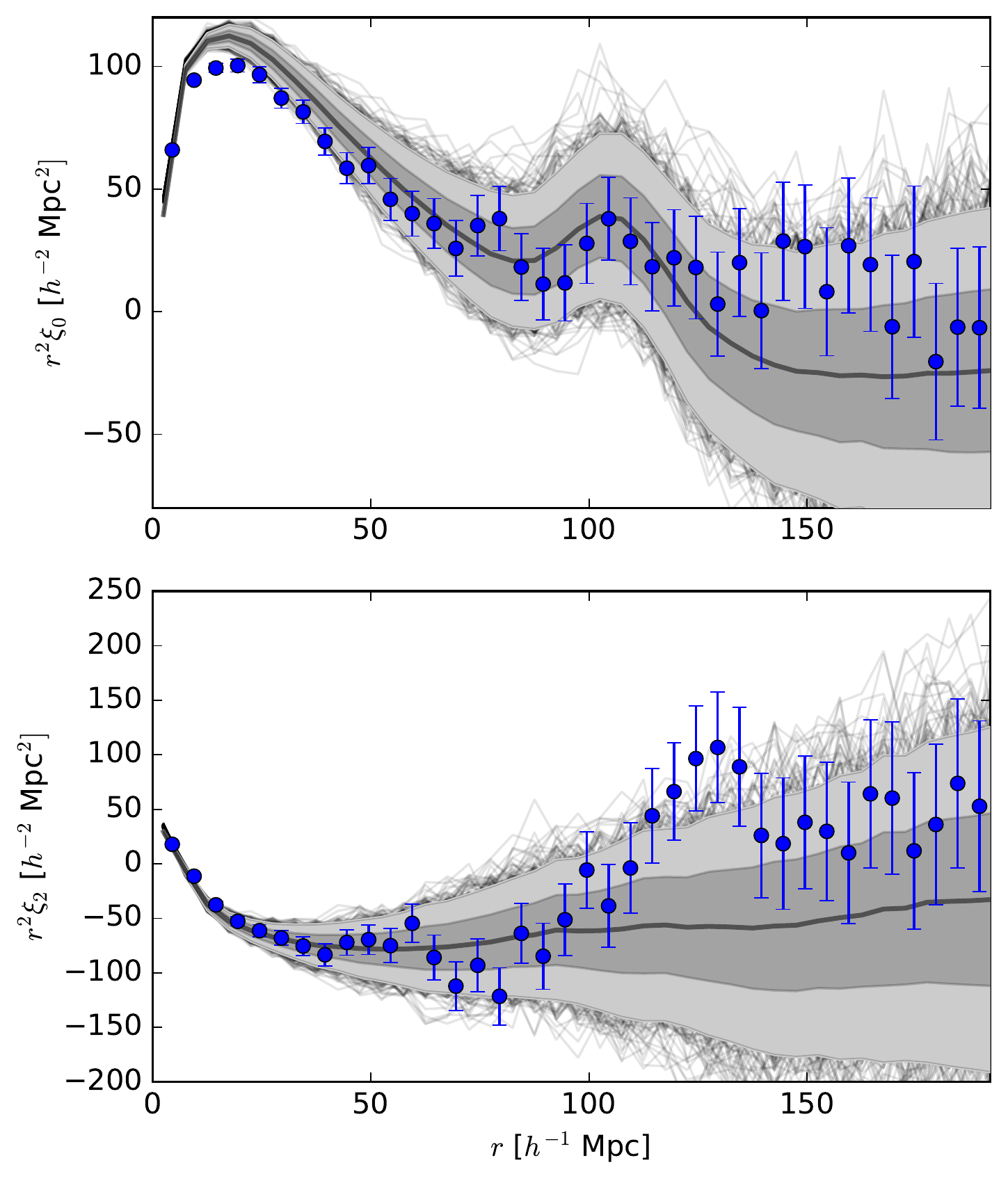}
\includegraphics[width=0.49\textwidth]
{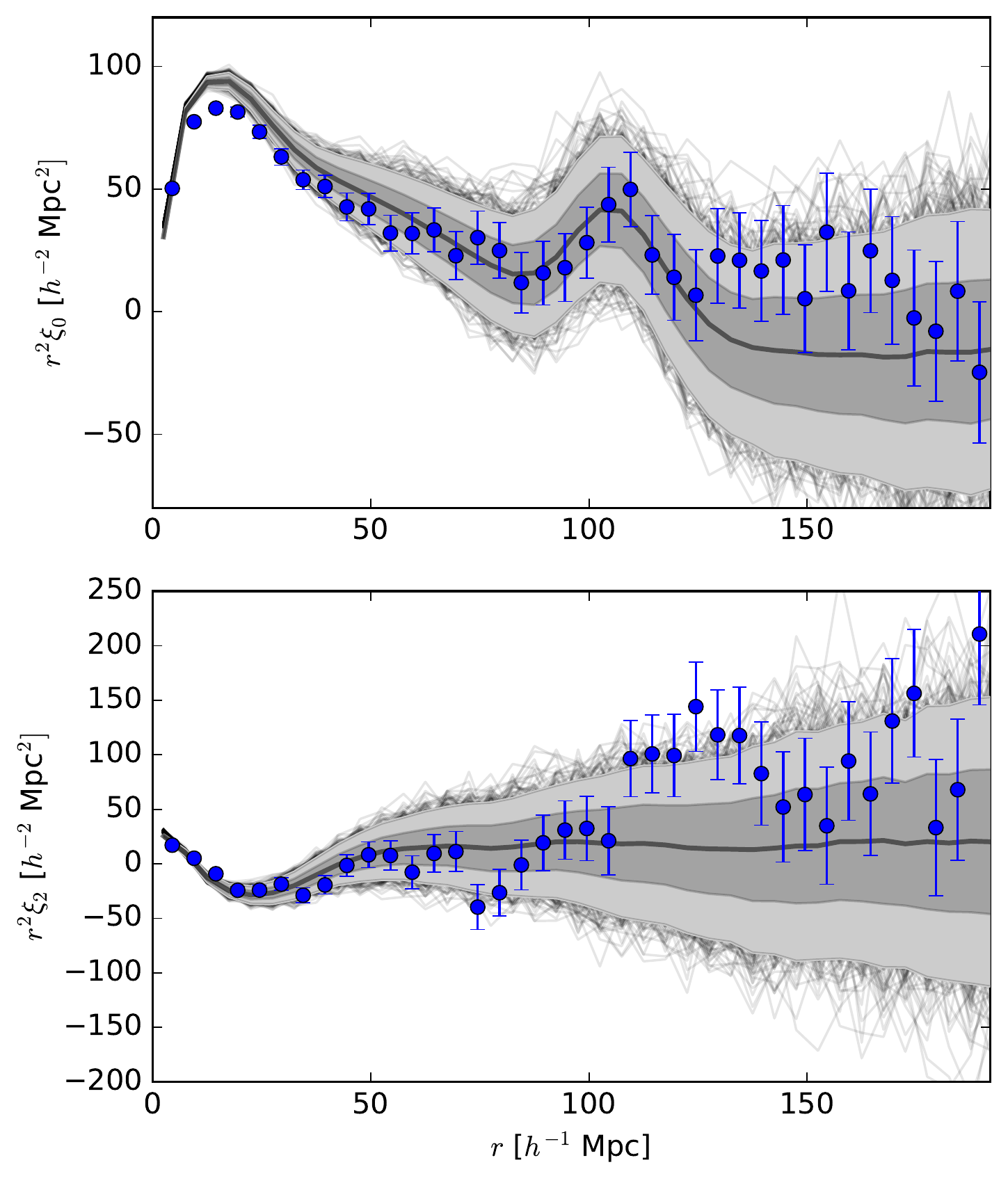}
\caption{ Comparison of the monopole (top panels) and quadrupole
  (bottom panels) of data (blue points) and mock catalogs (gray lines)
  of the eBOSS+CMASS combined NGC+SGC.  Panels on the left
  (right) show the pre(post)-reconstruction catalogs.
  The shaded regions represent the 68 and 95\% boundaries of the
  distributions of correlation functions around the mean. Mocks and data 
  correlation functions agree on most scales, except in the monopole at 
  $r<30$\hmpc\ (excluded from our analysis) and around $r\sim 120$\hmpc\
  in the quadrupole. }
\label{fig:multipoles-data-mocks}
\end{figure}

We computed the correlation function for all 1000 mock catalogs 
of the eBOSS+CMASS samples, pre and post reconstruction. 
Our fiducial cosmology was employed to compute comoving separations. 
The monopoles and quadrupoles for the mock catalogs are
displayed in Figure~\ref{fig:multipoles-data-mocks} and compared
with the data. For scales of interest for BAO
($r > 30$\hmpc) the mock and data results mostly agree, 
except on few points in the quadrupole where the data show deviations 
of about 3$\sigma$ 
from the mean of the mocks. The source of this
deviation was not identified, but broadband terms 
(Eq.~\ref{eq:template}) in our fits are able to marginalize over 
part of these deviations. 

\begin{figure}[t]
\centering
\includegraphics[width=0.32\textwidth]{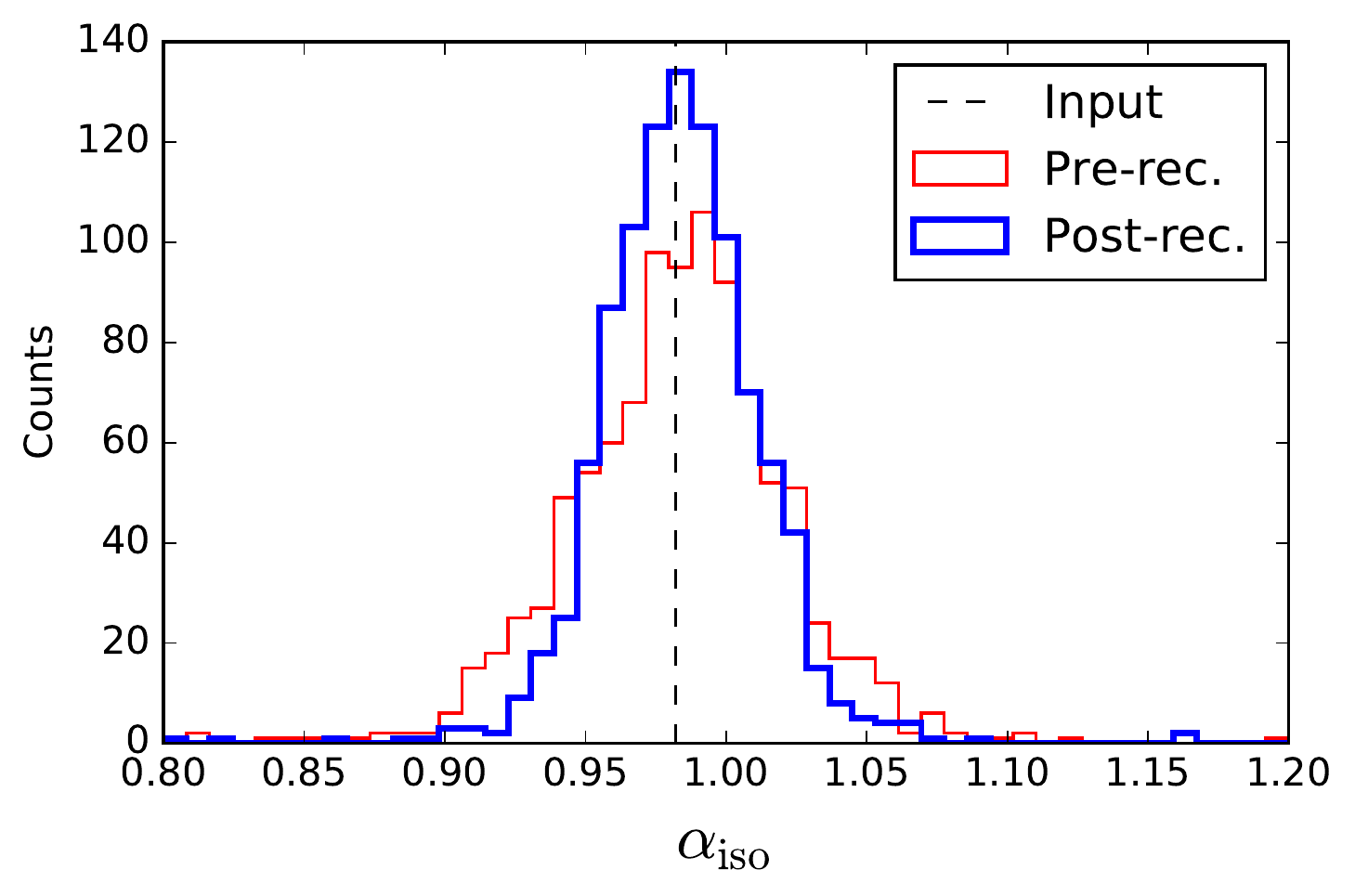}
\includegraphics[width=0.32\textwidth]{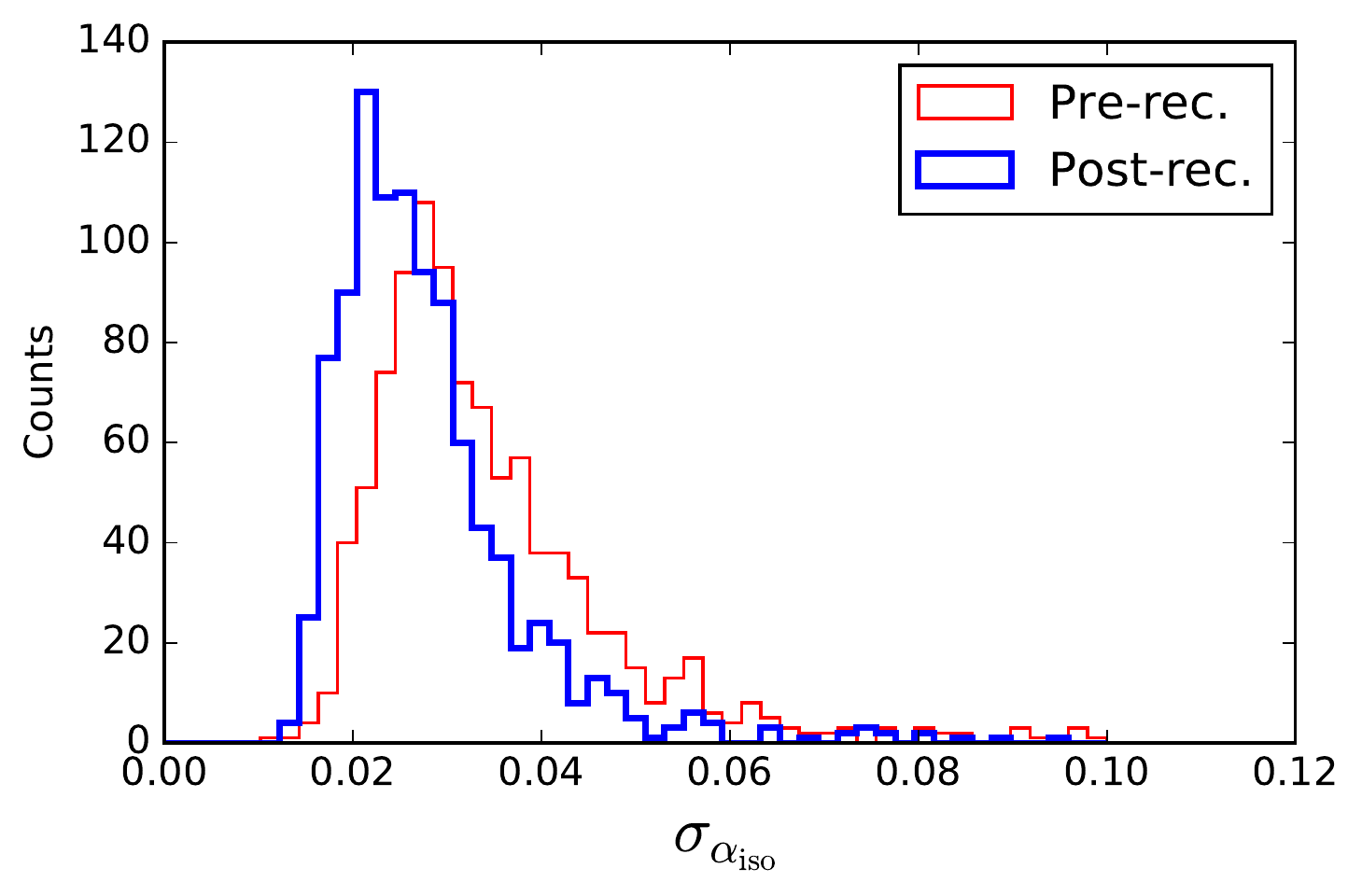}
\includegraphics[width=0.32\textwidth]{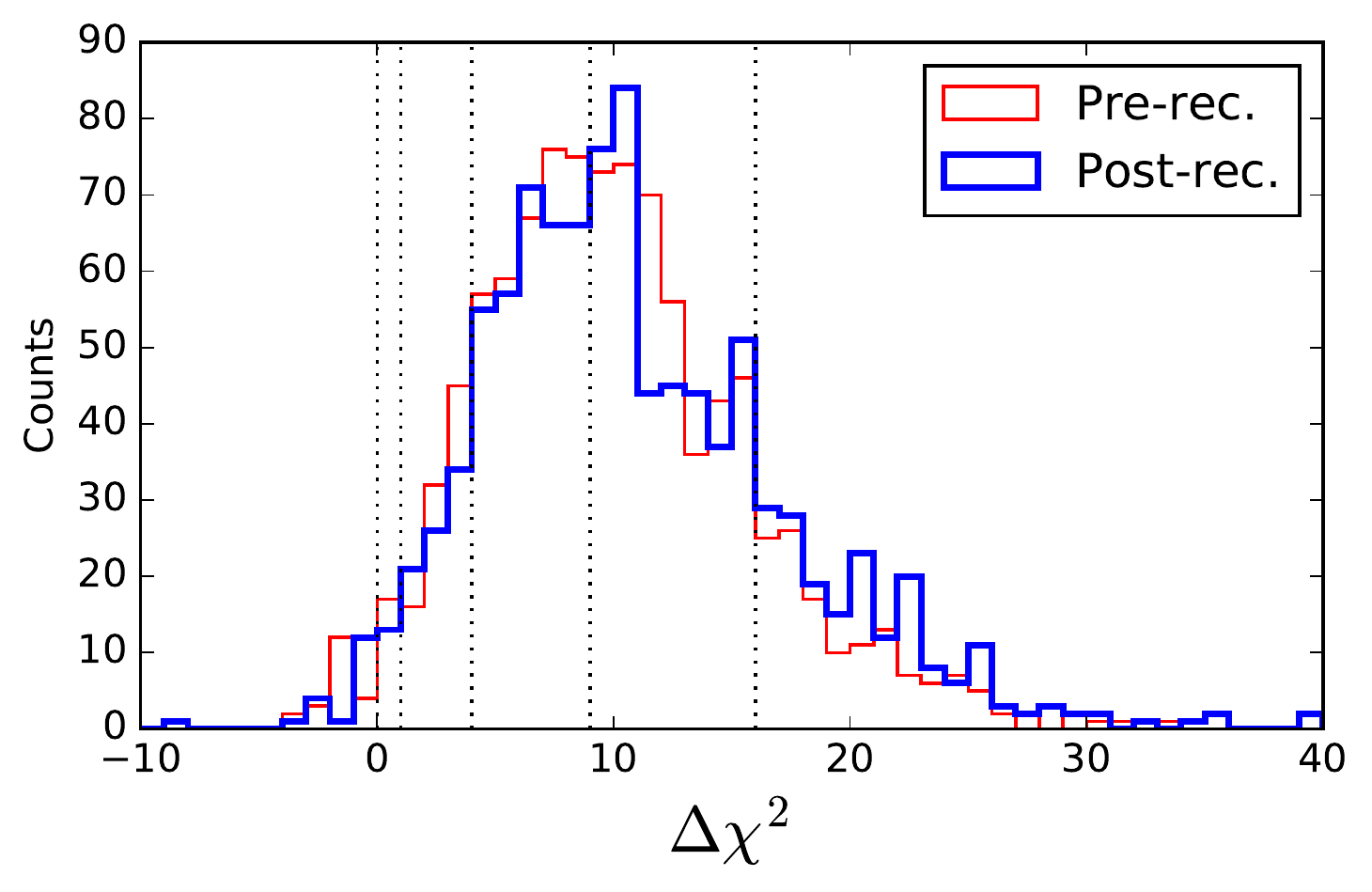}
\caption{The distribution of best-fit values for $\aiso$ (left), 
estimated errors (center), and the 
$\Delta \chi^2$ (right panel) for the set of 1000 mock catalogs. 
Results before reconstruction are presented in red while results
after reconstruction (method C) are presented in blue. 
Dashed lines on left panels
indicate the expected value given cosmological model in 
Table~\ref{tab:cosmologies}. 
Dotted lines in the right panel indicates
$\Delta \chi^2 = \{0, 1.0, 4.0, 9.0, 16.0\}$, corresponding to 
BAO detection significances of 0, 1, 2, 3 and 4$\sigma$, respectively. }
\label{fig:hists_iso}
\end{figure}

\begin{figure}[t]
\includegraphics[width=0.33\textwidth]
{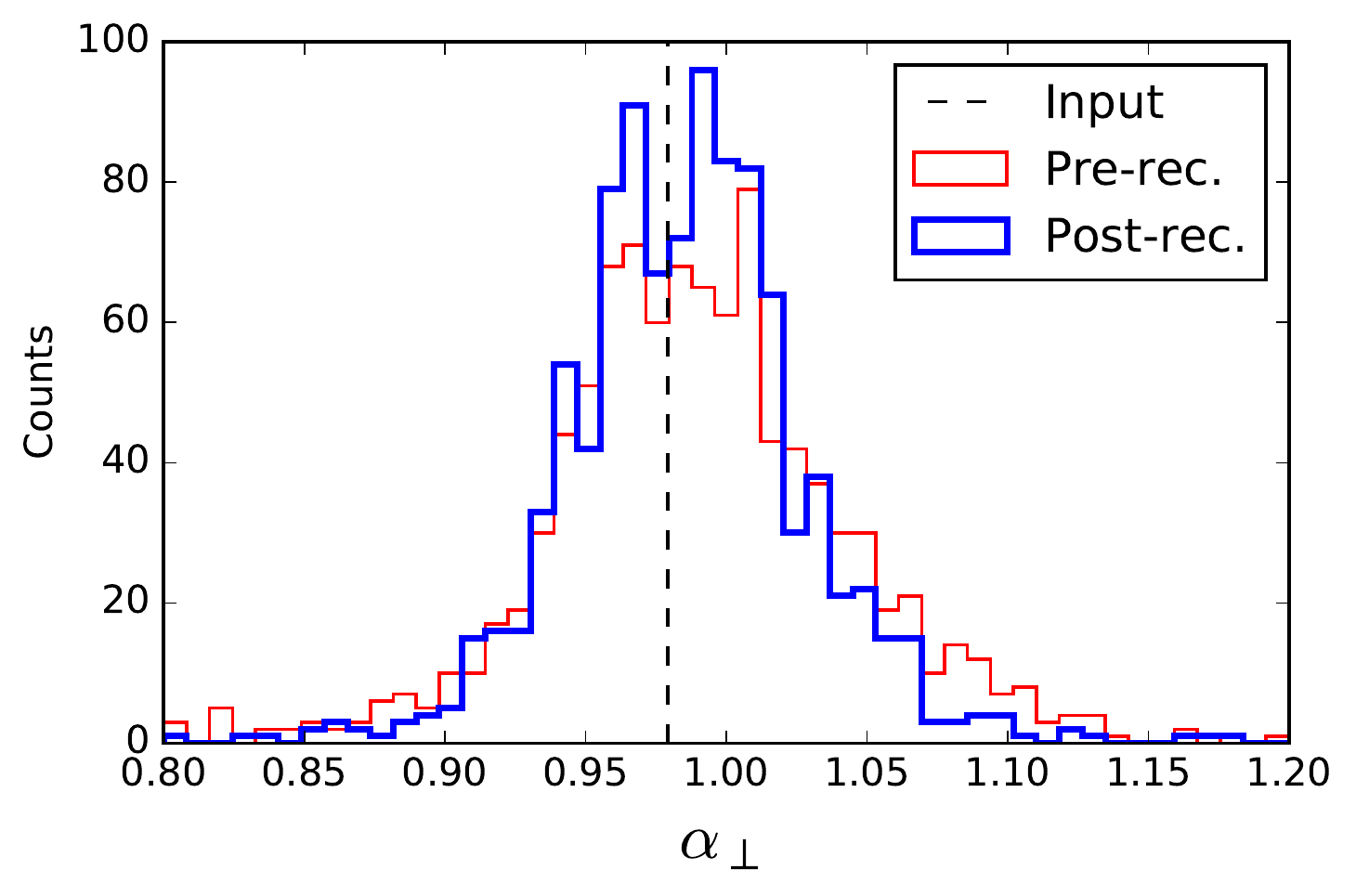}
\includegraphics[width=0.33\textwidth]
{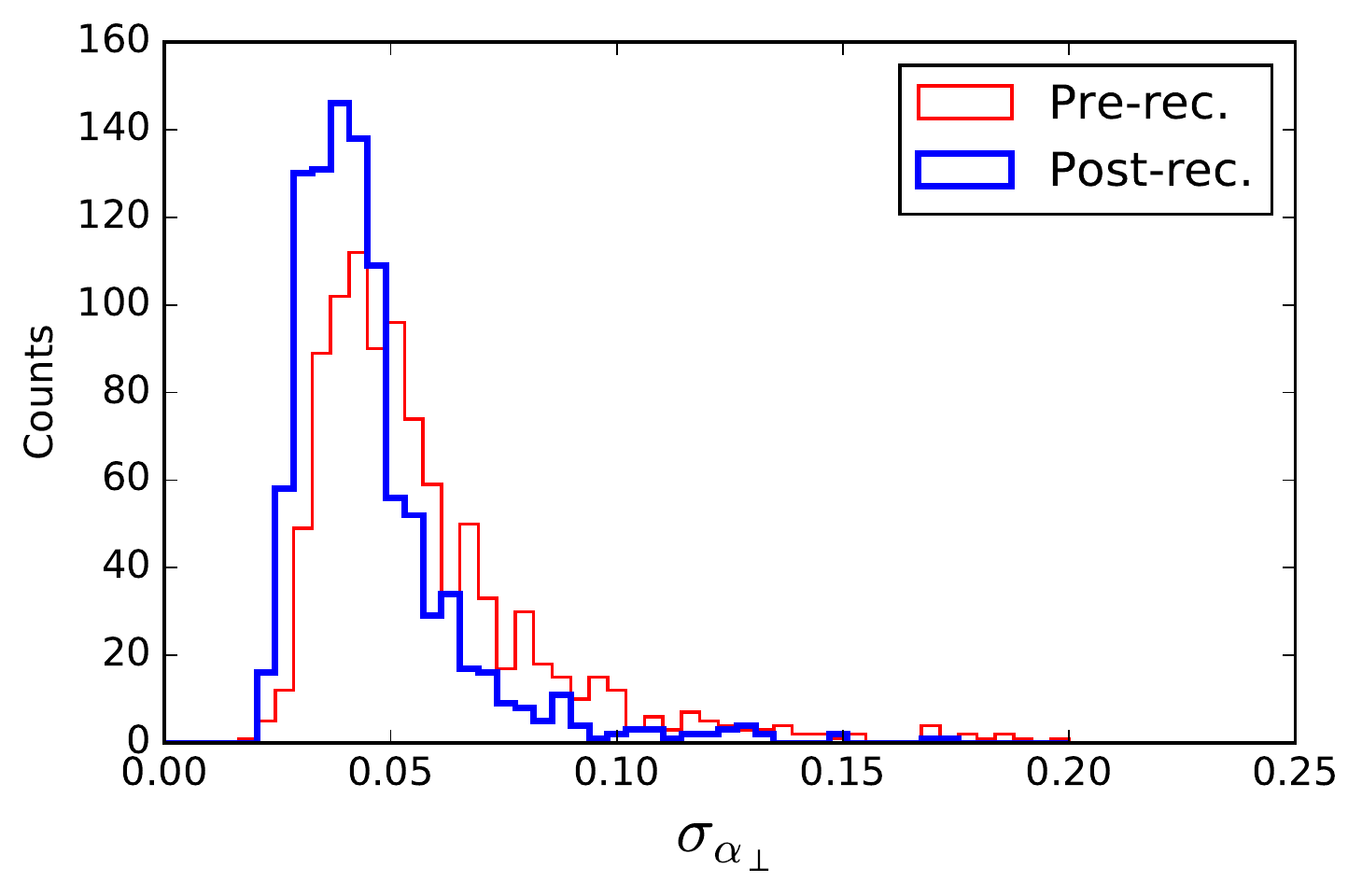}
\includegraphics[width=0.33\textwidth]
{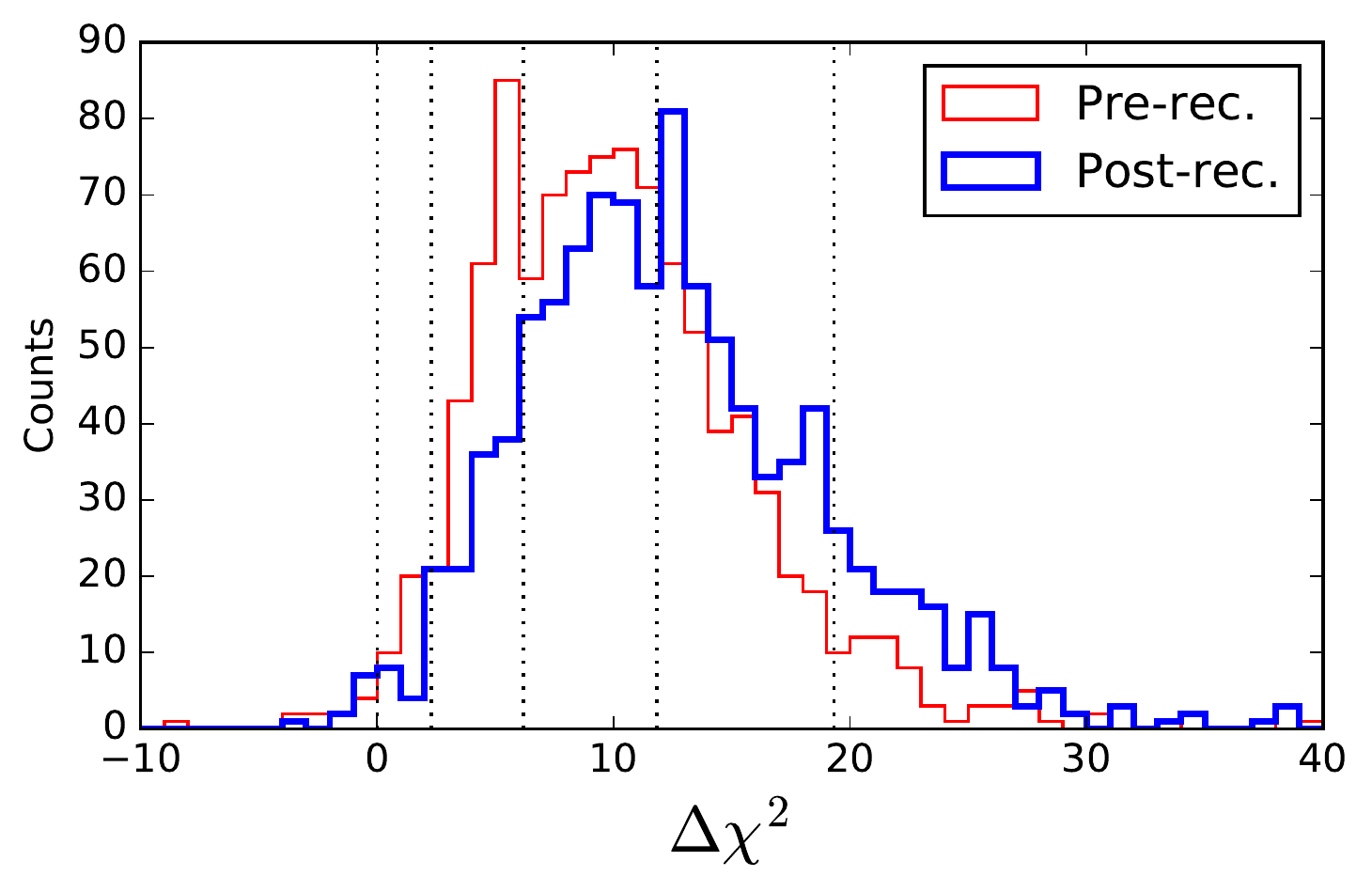}
\includegraphics[width=0.33\textwidth]
{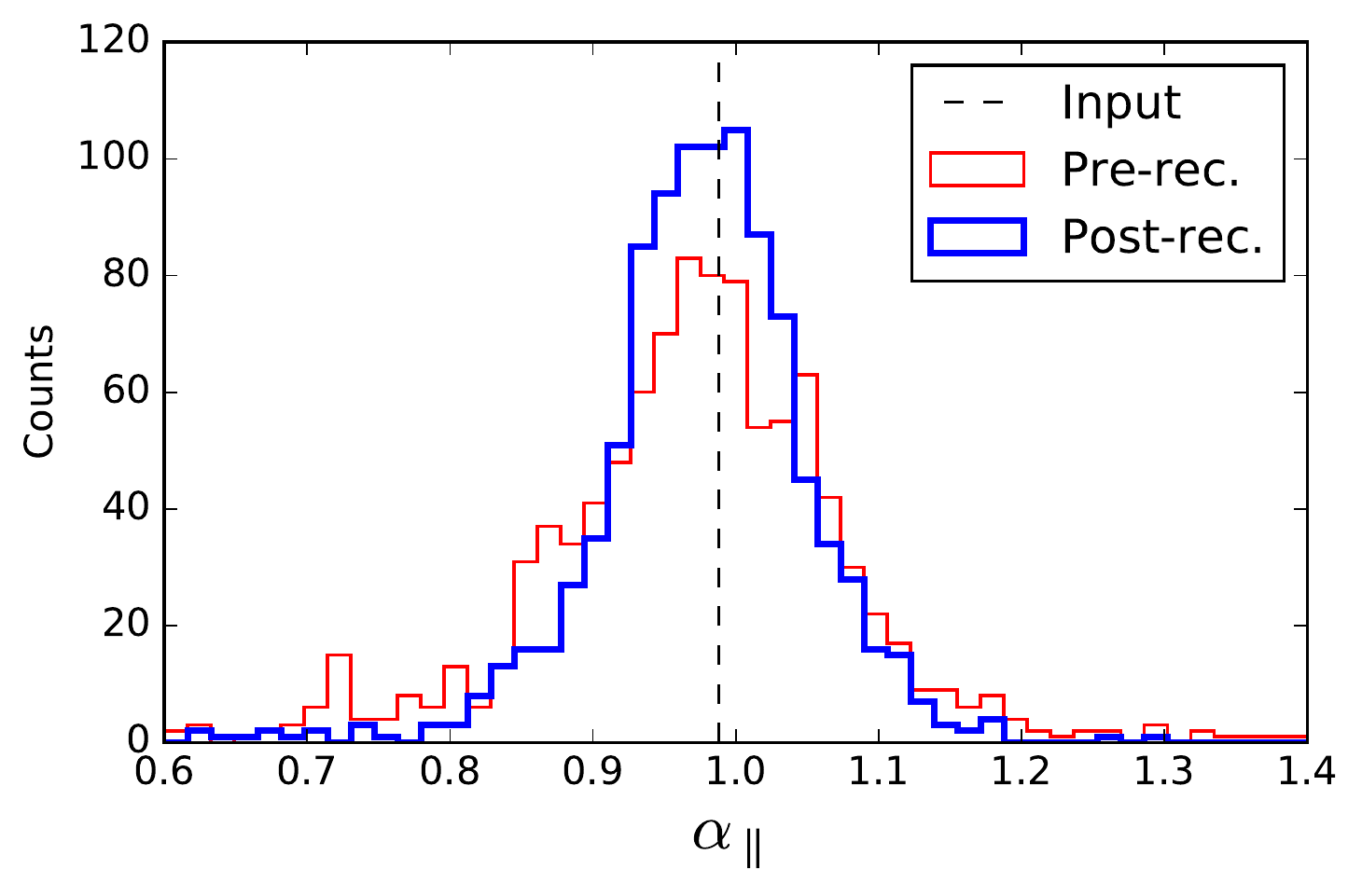}
\includegraphics[width=0.33\textwidth]
{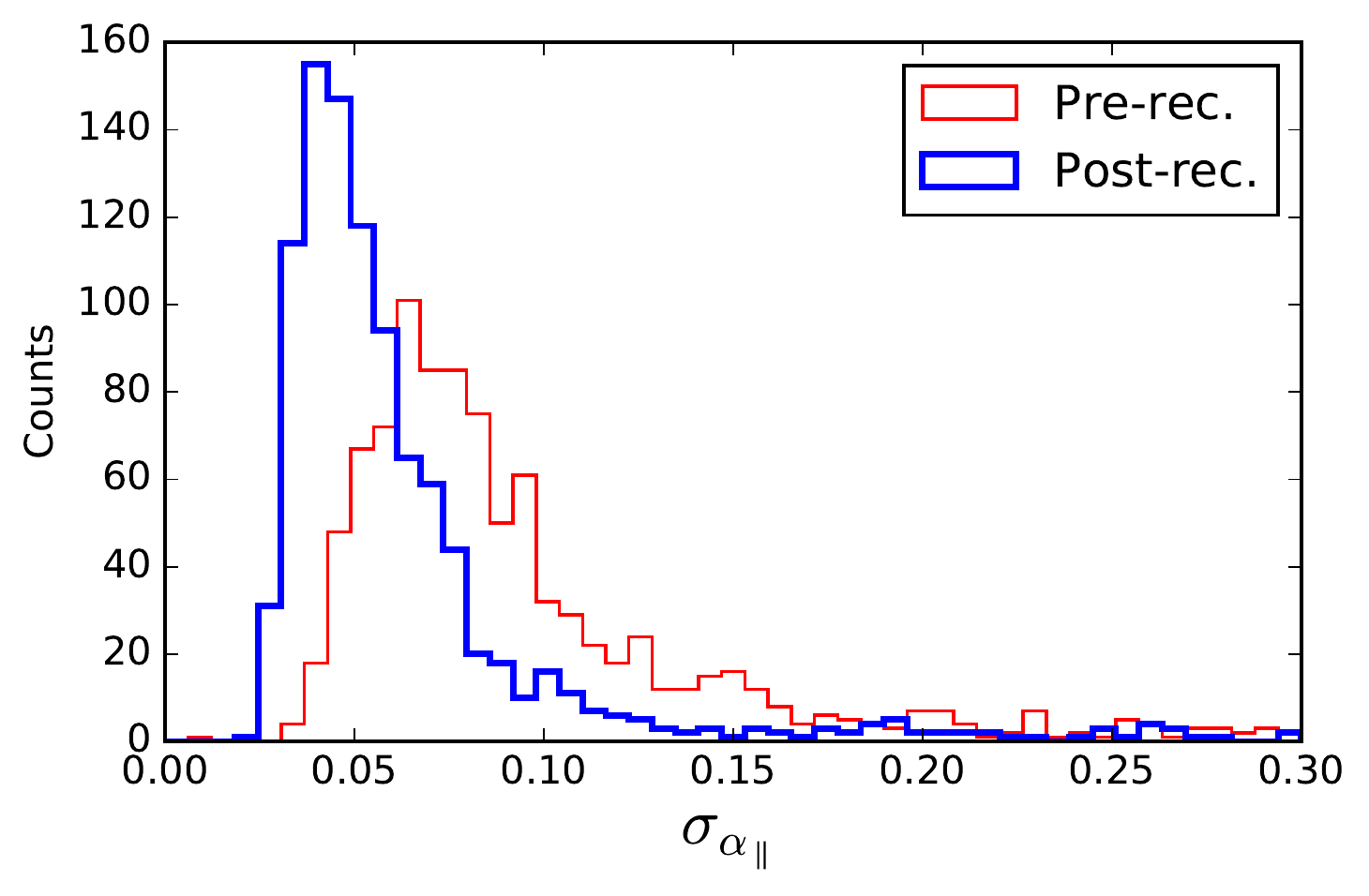}
\caption{
Same format as Fig.~\ref{fig:hists_iso} but for $\aperp$ and $\apar$.
Dotted lines in the right panel indicate
$\Delta \chi^2 = \{0, 2.30, 6.18, 11.83, 19.33\}$, corresponding to 
BAO detection significances of 0, 1, 2, 3 and 4$\sigma$, respectively.
The reduction of the width of the $\alpha$ histograms is also seen as 
a global reduction of $\sigma_{\alpha}$ in the middle panels. }
\label{fig:hists_aniso}
\end{figure}

We performed fits of the BAO scale over $30 < r< 180$~\hmpc\ 
on the set of 1000 mock catalogs
in order to check for any possible biases or mis-estimation of errors. 
Table~\ref{tab:mock_fits} and Fig.~\ref{fig:hists_iso} show results 
for isotropic fits while Table~\ref{tab:mock_fits_aniso} and 
Fig.~\ref{fig:hists_aniso} present anisotropic results. 
Dilation factors are compared to their expected values, which are
not unity given the different cosmological models used in the 
simulations and the analysis (see Table~\ref{tab:cosmologies}).

\begin{table}[t]
\centering
\caption{Results of isotropic fits on the full set of 1000 mock catalogs
of eBOSS+CMASS NGC+SGC. Given the cosmological parameters in Table~\ref{tab:cosmologies},
the input value of ${\aiso}_0$ is $0.9821$. All fits use $\Delta r = 5$\hmpc\ bins. 
The column $N_{good}$ shows the number of mocks where $\Delta \chi^2>1.0$ 
from which all numbers are computed.}
\label{tab:mock_fits}
\begin{tabular}{lcccccc}
\hline
\hline
case & $N_{good}$ & $\langle \aiso - {\aiso}_0 \rangle$ & rms$(\aiso)$ & $\langle \sigma_{\aiso} \rangle$  & $\langle \chi^2_r \rangle$ & $\langle \Delta \chi^2 \rangle$ \\
Pre-rec. & 961 & $0.0015 \pm 0.0012$ & 0.037 & 0.034 &  1.00 & 10.5  \\
Post-rec (A) & 968 & $0.0045 \pm 0.0009$ & 0.029 & 0.029 &  1.00 & 10.7  \\
Post-rec (B) & 960 & $-0.0012 \pm 0.0010$ & 0.032 & 0.029 &  1.00 & 10.8  \\
Post-rec (C) & 973 & $0.0001 \pm 0.0008$ & 0.026 & 0.026 &  0.99 & 12.1  \\
\hline
\hline
\end{tabular}
\end{table}

For the isotropic fits, values of $\aiso$ are consistent with the 
input within the 2$\sigma$ level for pre-reconstructed mocks and for 
two of the reconstruction methods (B and C). The rms of $\aiso$ 
values for pre-reconstructed mocks are larger than the average
per-mock estimated error due to outliers of $\aiso$. 
For the post-reconstruction mocks, the rms and mean error are 
in good agreement. The average gain in estimated errors caused by 
reconstruction is $\sim20$\% and is clearly visible in 
Fig.~\ref{fig:hists_iso}. The significance of BAO detections
are estimated in individual mocks by comparing the $\chi^2$ 
of a model with peak to the $\chi^2_{\rm no~peak}$ of a model 
without peak: $\Delta \chi^2 = \chi^2_{\rm no~peak} - \chi^2$.
Pre-reconstruction mocks show $\langle \Delta \chi^2 \rangle=10.5$
corresponding to $\sim 3.2\sigma$ detection.   
Reconstruction slightly increases the significance by 0.2 to 1.6 unit 
in $\Delta \chi^2$, depending on the reconstruction procedure, 
reaching 3.5$\sigma$ for case C. 
Less than 4\% of mocks produce no 
BAO detection, where $\Delta \chi^2 < 1$, even after 
reconstruction is applied. 

Anisotropic fits on mock catalogs have similar behavior to the 
isotropic ones. Figure~\ref{fig:hists_aniso} reveals 
slightly more Gaussian distributions of $\aperp$ and 
$\apar$ after applying reconstruction.
However, the mean values of $\alpha$ show a small bias relative to
the expected input values. These biases represent 
20\% of the expected error of a single realization for 
pre-reconstruction mocks and they are reduced to  5-15\% in 
post-reconstruction cases. These biases are caused mostly by
the low statistical power of the current LRG sample which makes
the $\alpha$ distributions non-Gaussian. 
If we fit the average correlation function of 1000 mock catalogs 
we are able to recover the input values within error bars. 
The average $\Delta \chi^2$ is virtually the same as the 
isotropic fits for both pre and post-reconstruction cases. 
Since we have two BAO parameters instead of one, 
the corresponding significance post-reconstruction is $\sim 3\sigma$. 
If mock catalogs are a good representation 
of real data, we expect a $\sim$4.8\% measurement on $\aperp$ and 
$\sim 8.5$\% on $\apar$ post-reconstruction.

\begin{table}[t]
\centering
\caption{Results of anisotropic fits on the full set 
of 1000 reconstructed mock catalogs. 
Given the cosmological parameters in Table~\ref{tab:cosmologies},  the 
input values are: ${\aperp}_0 = 0.9791$ and ${\apar}_0 = 0.9880$. All fits are using 
$\Delta r = 5$\hmpc\ bins. 
The column $N_{good}$ shows the number of mocks where $\Delta \chi^2>2.3$ 
from which all numbers are computed.}
\label{tab:mock_fits_aniso}
\begin{tabular}{lccccccccc}
\hline
\hline
case & $N_{good}$ & $\langle \aperp - {\aperp}_0 \rangle$ & rms$(\aperp)$ & 
       $\langle \sigma_{\aperp} \rangle$ &  
       $\langle \apar - {\apar}_0\rangle$ & rms$(\apar)$ & 
       $\langle \sigma_{\apar} \rangle$ &  
       $\langle \chi^2_r \rangle$ &  
       $\langle \Delta \chi^2 \rangle$ \\
Pre-rec            & 926 & $0.0120 \pm 0.0021$ & 0.063 & 0.057 & $-0.0192 \pm 0.0037$ & 0.113 & 0.101 & 0.96 & 10.7 \\
Post-rec (A)  & 938 & $0.0078 \pm 0.0016$ & 0.048 & 0.047 & $0.0090 \pm 0.0027$ & 0.084 & 0.071 & 0.98 & 12.1 \\
Post-rec (B) & 938 & $0.0030 \pm 0.0016$ & 0.050 & 0.048 & $-0.0011 \pm 0.0029$ & 0.089 & 0.088 & 0.97 & 11.9 \\
Post-rec (C)  & 951 & $0.0072 \pm 0.0014$ & 0.044 & 0.044 & $-0.0118 \pm 0.0026$ & 0.080 & 0.062 & 0.96 & 12.9 \\
\hline
\hline
\end{tabular}
\end{table}

\section{Results}
\label{sec:results}

In this section, we present our measurements, and the basic tests
performed on the robustness of those measurements.

\subsection{Fits to the Data}

We performed BAO fits on the final eBOSS+CMASS NGC+SGC sample 
(unless otherwise stated), using the covariance matrix derived 
from mock catalogs. We scale our covariance matrices by a factor of
0.9753 to account for the slight mismatch in footprint area between 
data and mocks. In order to test the robustness of our measurements, 
fits were performed with a set of different choices for binning,
$\Delta r$, and
 separation ranges, $r_{\rm min} < r < r_{\rm max}$. Our fiducial 
reconstruction method is method C (see section~\ref{sec:recon}). 

\begin{figure}[t]
\centering
\includegraphics[width=0.4\textwidth]
{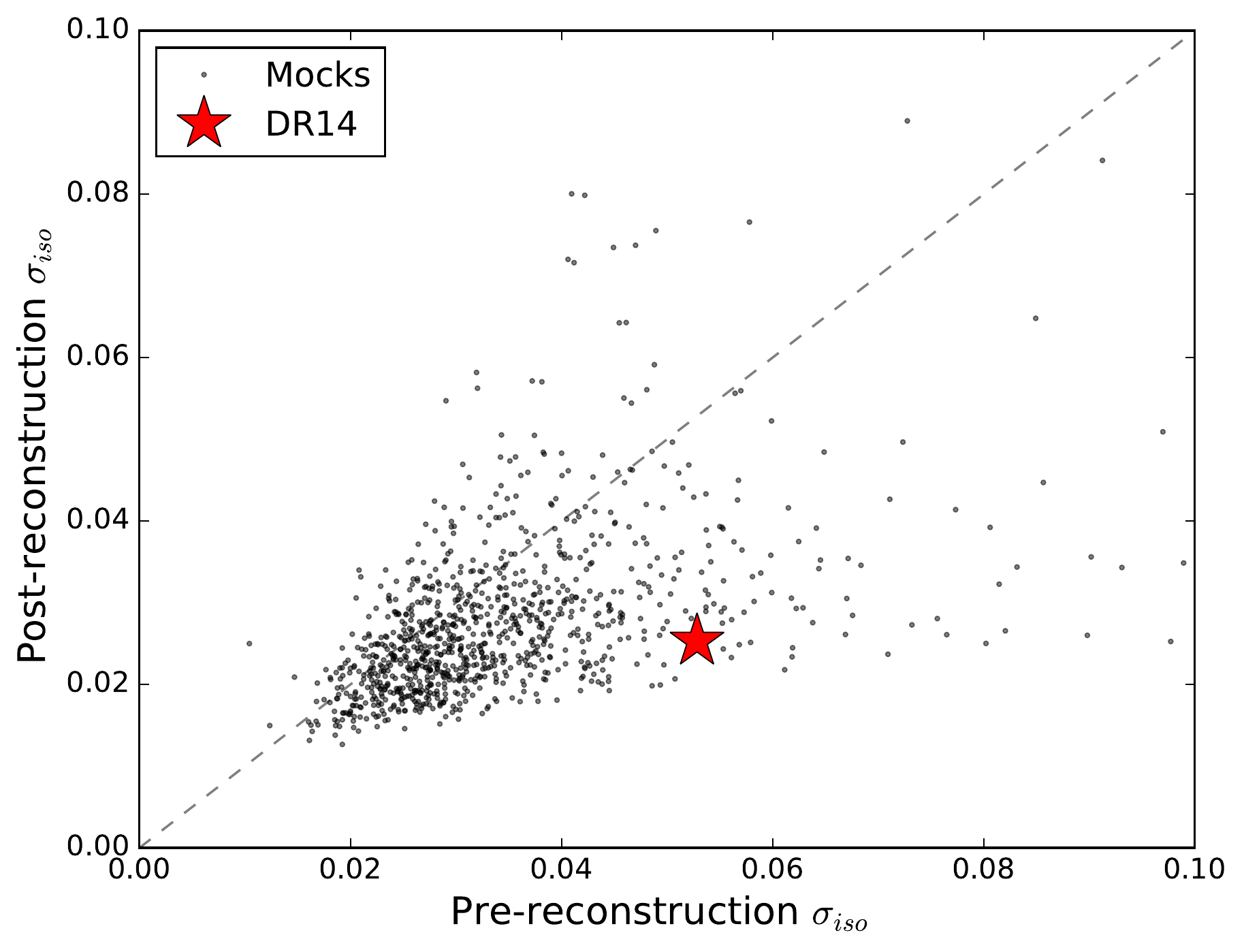}
\caption{Estimated errors for the 1000 mocks. Pre-reconstruction 
results are shown in the x-axis and the post-reconstruction results (method C) 
are plotted on the y-axis. The red star indicates the result with our data 
sample. While the error pre-reconstruction with data is quite large  
compared to mocks, the error post-reconstruction is typical of those
found in mocks.  
}
\label{fig:scatter-data-mocks}
\end{figure}

Table~\ref{tab:data_fits} presents the results for isotropic fits pre and 
post-reconstruction. All $\aiso$ values are stable and 
differ by less than 1$\sigma$. Estimated errors are also stable and 
no trend is seen when varying 
the choice of binning and fitting range.
The average significance for BAO detection in this sample 
is $\sim 1.5\sigma$ for pre-reconstruction, increasing to 
$\sim 2.7\sigma$ after reconstruction.
The estimated error on $\aiso$ is reduced by $\sim 40$\% on average
and $\chi^2$ values become closer to the number of degrees
of freedom after reconstruction.
The observed reduction in errors 
represents a significant improvement when compared to mock catalogs. 
Fig.~\ref{fig:scatter-data-mocks} compares $\sigma_{\aiso}$ obtained
pre and post-reconstruction for mocks and data. 
While the data appear to be at the extremes of the distribution, 
the error post-reconstruction is typical of that found in mocks. 

\begin{table}[t]
\centering
\caption{Isotropic fits on data for different choices of binning.  
In bold are the results for our fiducial analysis.}
\label{tab:data_fits}
\begin{tabular}{cccc|ccc|ccc}
\hline
\hline
\multicolumn{4}{c}{case} & \multicolumn{3}{c|}{  pre-reconstruction } & \multicolumn{3}{c}{ post-reconstruction } \\
Sample & $\Delta r$ & $r_{\rm min}$ & $r_{\rm max}$ & $\aiso$ & $\chi^2/$dof & $\Delta \chi^2$ &  $\aiso$ & 
 $\chi^2/$dof & $\Delta \chi^2$\\
\hline

NGC+SGC & \bf{5} & \bf{32} & \bf{182} & $\bf{0.949^{+0.054}_{-0.051}}$ & \bf{28.6/25}         & \bf{2.1}            & $\bf{0.968^{+0.026}_{-0.025}}$ & \bf{18.9/25}         & \bf{7.8}         \\
        &  & 29 & 179 & $0.928^{+0.049}_{-0.040}$ & 45.4/25         & 3.0            & $0.975^{+0.029}_{-0.027}$ & 26.8/25         & 6.8         \\
        &  & 30 & 180 & $0.951^{+0.048}_{-0.049}$ & 51.5/25         & 1.8            & $0.978^{+0.025}_{-0.022}$ & 30.0/25         & 8.7         \\
        &  & 31 & 181 & $0.949^{+0.045}_{-0.040}$ & 33.8/25         & 3.1            & $0.966^{+0.026}_{-0.026}$ & 19.8/25         & 6.8         \\
        &  & 28 & 178 & $0.935^{+0.051}_{-0.045}$ & 33.0/25         & 2.6            & $0.966^{+0.027}_{-0.023}$ & 27.2/25         & 8.8         \\ 
        & 8 & 26 & 178 & $0.961^{+0.048}_{-0.047}$ & 17.2/13         & 2.2            & $0.974^{+0.024}_{-0.024}$ & 13.9/13         & 8.3         \\
        &  & 28 & 180 & $0.941^{+0.057}_{-0.045}$ & 9.5/14         & 2.3            & $0.958^{+0.029}_{-0.026}$ & 7.2/14         & 7.1         \\
        &  & 30 & 182 & $0.932^{+0.055}_{-0.047}$ & 15.3/14         & 2.4            & $0.968^{+0.025}_{-0.024}$ & 8.5/14         & 7.5         \\
        &  & 32 & 184 & $0.948^{+0.047}_{-0.046}$ & 17.2/13         & 2.9            & $0.987^{+0.028}_{-0.026}$ & 14.3/13         & 6.5         \\
NGC & 5 & 30 & 180 & $0.996^{+0.467}_{-0.079}$ & 49.4/25         & 0.0            & $0.987^{+0.032}_{-0.031}$ & 32.8/25         & 6.2         \\
SGC & 5 & 30 & 180 & $0.942^{+0.051}_{-0.049}$ & 59.5/25         & 2.3            & $0.953^{+0.047}_{-0.052}$ & 23.2/25         & 1.9         \\

\hline
\hline
\end{tabular}
\end{table}
 
Given the scatter in the significance of the measurement depending
on the choice of analysis, we define our fiducial analysis as 
the one producing the $\Delta \chi^2$ that is the closest to the average 
values of all cases in Table~\ref{tab:data_fits}, i.e., 
$\Delta r~=~5$~\hmpc\ and $(r_{\rm min}, r_{\rm max}) = (32, 182)$~\hmpc.
 We also adopt this
 choice of fiducial analysis in our anisotropic fits below. 

\begin{figure}[t]
\centering
\includegraphics[width=0.49\textwidth]
{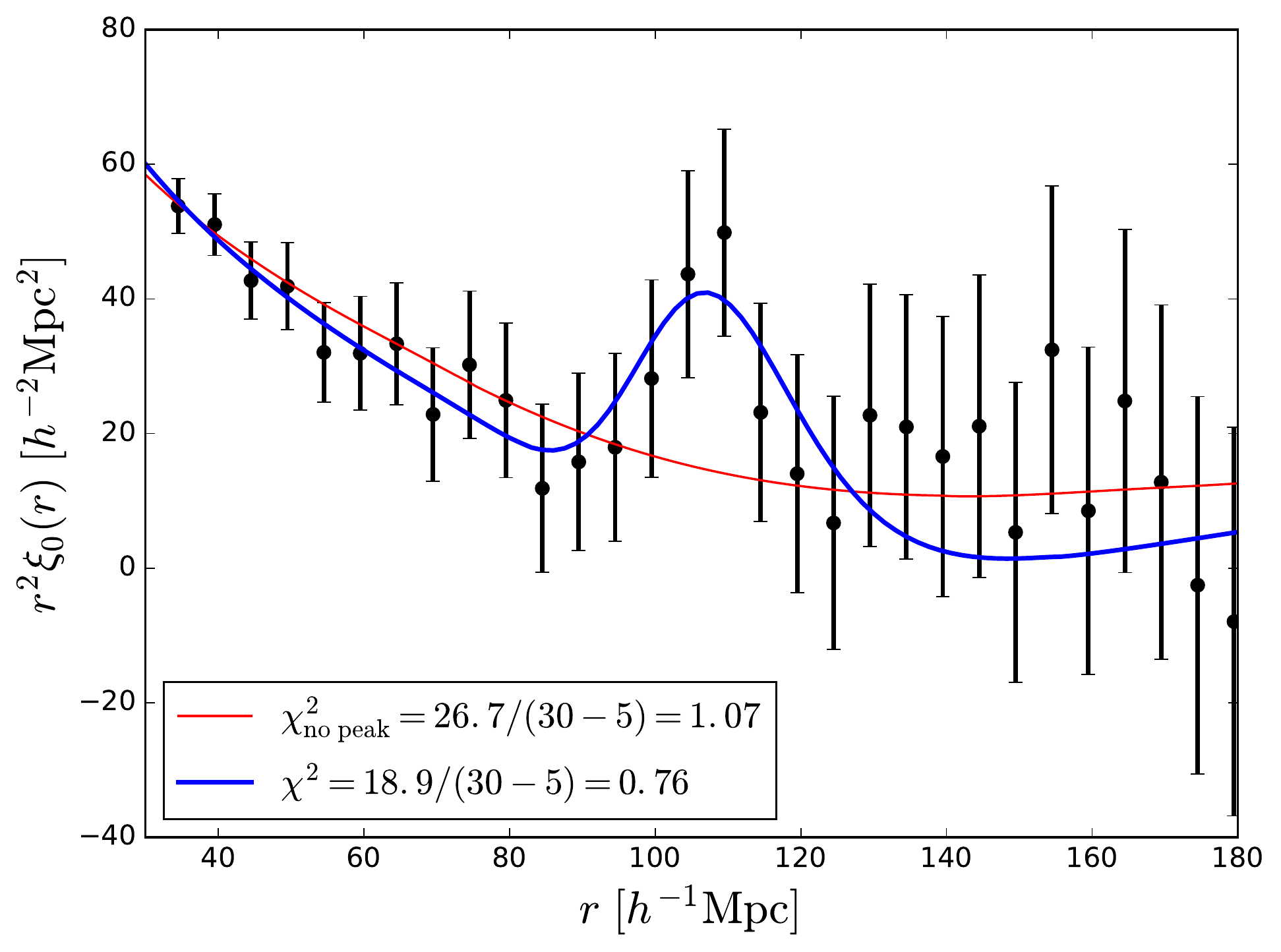}
\caption{Monopole of the post-reconstruction  
correlation function with the best-fit models with a BAO peak component 
(blue thick line) and without peak (red thin line) for our fiducial 
choice of binning ($\Delta r = 5$\hmpc\ and $32<r<182$\hmpc). }
\label{fig:monopole}
\end{figure}

\begin{figure}[t]
\centering
\includegraphics[width=0.45\textwidth]
{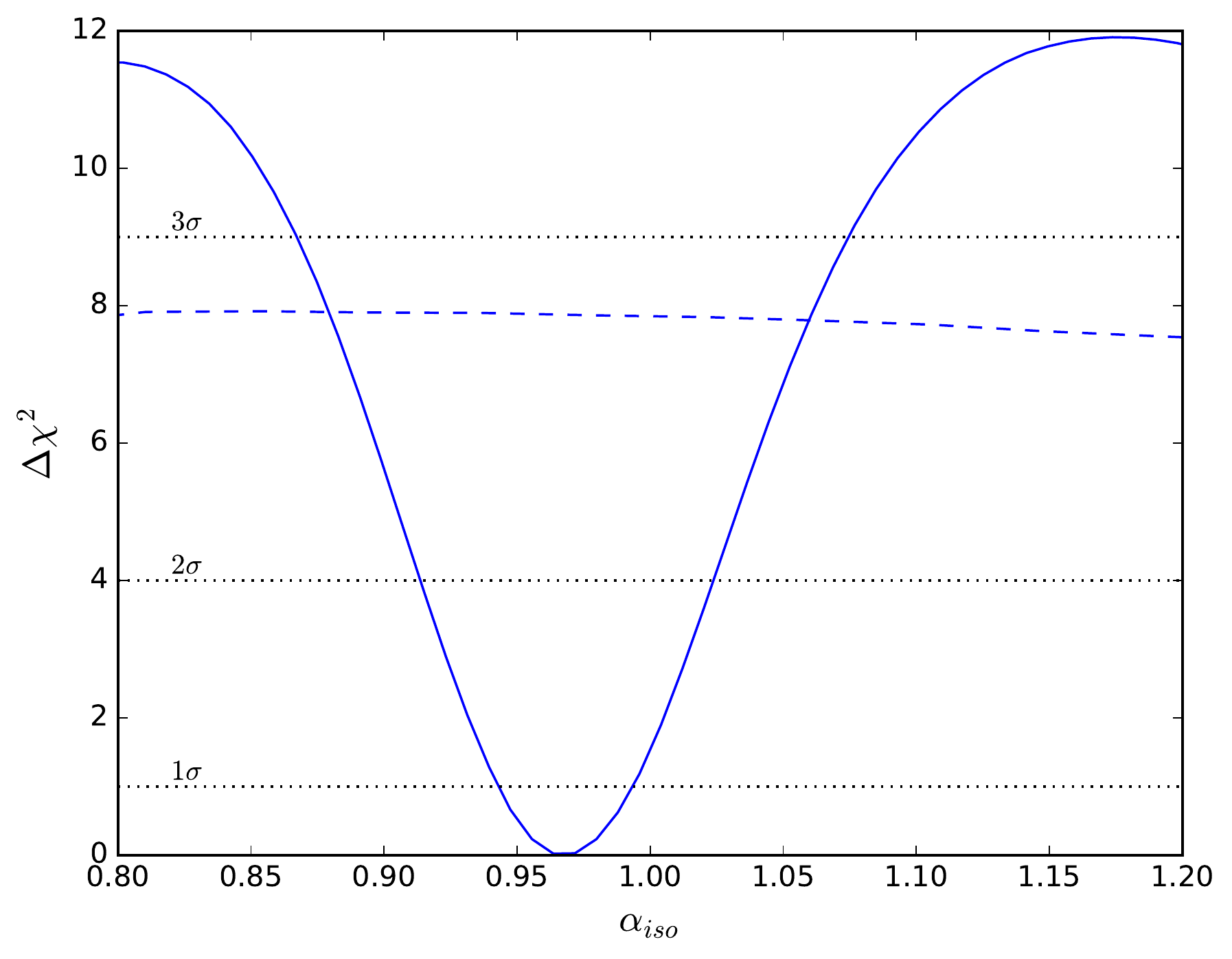}
\caption{$\chi^2$ as a 
function of the dilation parameter $\alpha_{\rm iso}$ for both models 
(solid: with peak, dashed: without peak). This curve corresponds to our fiducial choice of analysis 
($\Delta s = 5$\hmpc\ and $32< r< 182$\hmpc). The $\Delta \chi^2 = 7.8$ 
corresponds to a detection of 2.8$\sigma$ of the BAO peak. } 
\label{fig:chi2scan1d}
\end{figure}

Fig.~\ref{fig:monopole} shows the monopole and the two best-fit models 
(with and without BAO peak) on the reconstructed 
sample for our fiducial choice of analysis.  
Fig.~\ref{fig:chi2scan1d} presents the $\chi^2$ values as a function of 
$\aiso$ and marginalizing over all other parameters for both models. 
The model without peak has a $\chi_{\rm no \ peak}^2$ about 7.8 units
above the minimum of the model with peak, corresponding to a preference 
for the BAO peak model with a significance of 2.8$\sigma$.

Anisotropic fits are listed in Table~\ref{tab:data_fits_aniso} for 
the same analysis cases presented with isotropic fits. 
Best-fit $\aperp$ and $\apar$ values are mostly stable to 
changes in the analysis. However, errors on the pre-reconstruction
sample are quite unstable due to the low significance of the 
measurement ($\Delta \chi^2 \sim 3$ in average). Reconstruction  
stabilizes errors and increases the significance of our 
constraints ($\Delta \chi^2 \sim 8$ in average). The estimated errors
post-reconstruction are similar to average values
found in mock catalogs (c.f. Table~\ref{tab:mock_fits_aniso}). 

\begin{figure}[t]
\centering
\includegraphics[width=0.4\textwidth]
{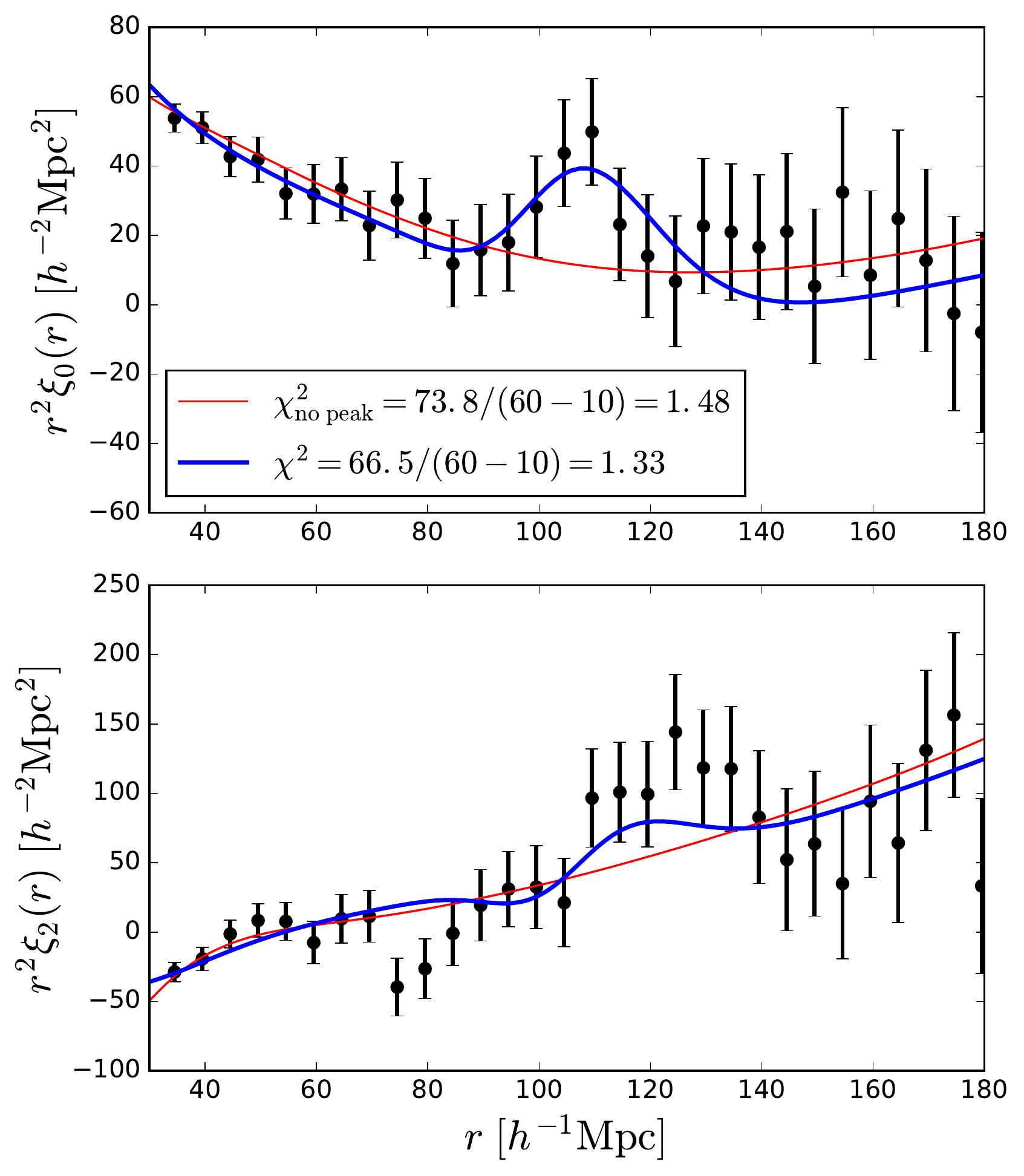}
\includegraphics[width=0.59\textwidth]
{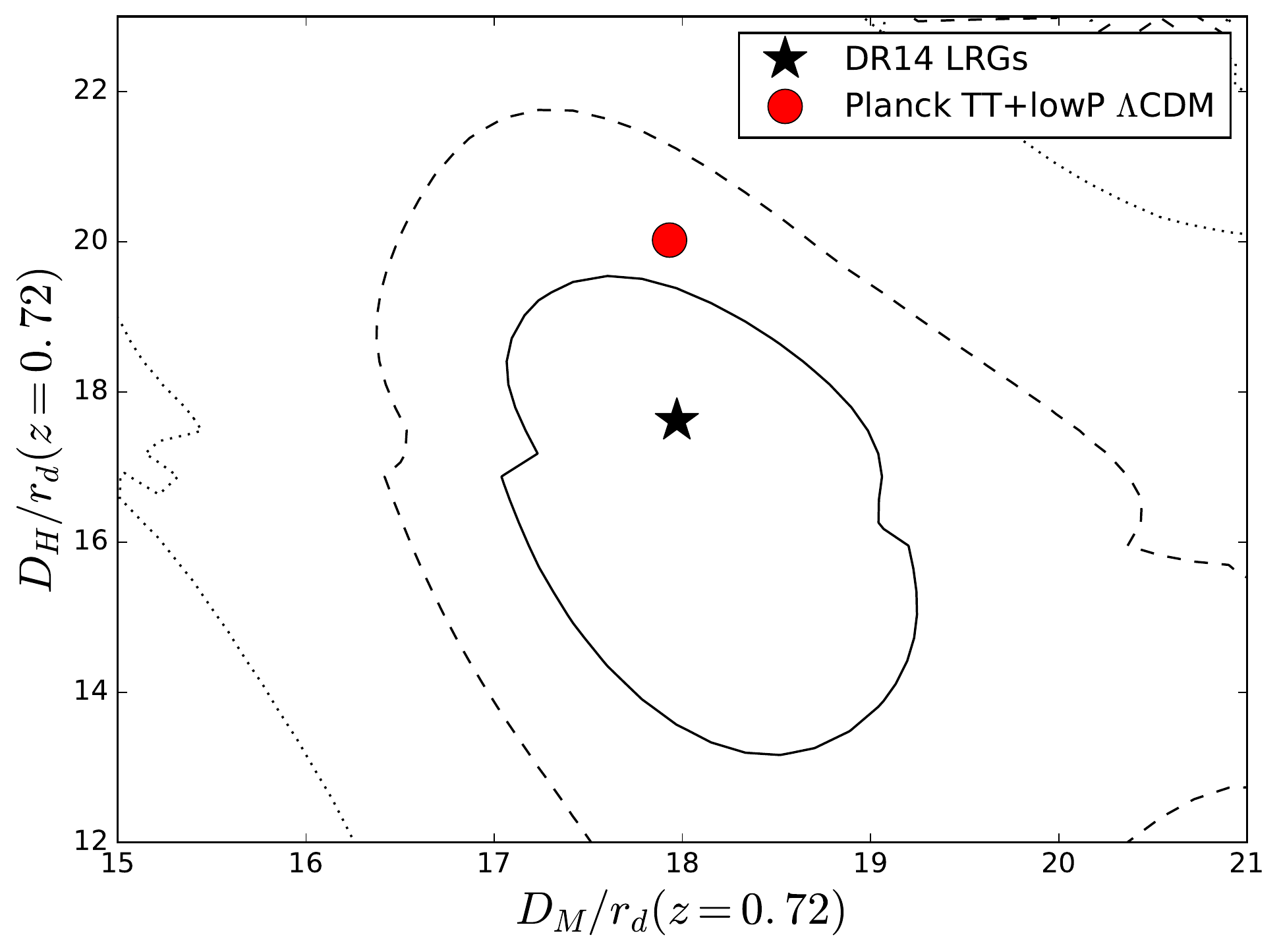}
\caption{ 
Left panel: monopole (top) and 
quadrupole (bottom) of the post-reconstruction correlation function 
with the best-fit 
models with (blue thick line) and without (red thin
line) a BAO peak component. Right panel: $\chi^2$ contours as a function of the dilation 
parameters $\alpha_\perp, \alpha_\parallel$ converted to $D_M/r_d$ and 
$D_H/r_d$ using our fiducial cosmology in Table~\ref{tab:cosmologies}.
The contours correspond to 
$\chi^2-\chi^2_{\rm min} = 2.3, 6.18, 11.83$ (solid, dashed and dotted,
respectively). For this fit, we obtain $\Delta \chi^2 = 7.2$. }
\label{fig:2dfits}
\end{figure}

The left panel of Fig.~\ref{fig:2dfits} displays the best-fit anisotropic 
 models compared to the data for our fiducial choice of analysis. 
The right panel shows the 
two dimensional $\chi^2$ contours for 1, 2 and 3$\sigma$, after
converting $\aperp$ and $\apar$ into $D_M/r_d$ and $D_H/r_d$ 
respectively, using values of our fiducial cosmological model on
Eq.~\ref{eq:aperp_apar}. The likelihood becomes highly 
non-Gaussian beyond the $\Delta \chi^2 = 2.3$ contour due to the low statistical
power of this sample. We expect that these contours will become
more Gaussian as we increase the size our data sample in the 
future. 

\begin{table}[t]
\centering
\caption{Anisotropic fits on data for different binning choices. In bold are
the results for our fiducial analysis (see text). Given the low statistical
power of this sample for anisotropic fits, we do not include separate 
results for NGC or SGC.}
\label{tab:data_fits_aniso}
\begin{tabular}{ccc|ccccc|ccccc}
\hline
\hline
\multicolumn{3}{c}{case} & \multicolumn{5}{c|}{pre-reconstruction} & \multicolumn{5}{c}{post-reconstruction} \\
 $\Delta r$ & $r_{\rm min}$ & $r_{\rm max}$ & 
$\alpha_\perp$ & $\alpha_\parallel$ & $\chi^2/$dof & $\Delta \chi^2$ & corr. & 
$\alpha_\perp$ & $\alpha_\parallel$ &  $\chi^2/$dof & $\Delta \chi^2$ & corr. \\
\hline

 \bf 5 & \bf 32 & \bf 182 &  $\bf 0.99^{+0.08}_{-0.05}$ & $\bf 0.82^{+0.09}_{-0.08}$ & \bf 75.2/50         & \bf 1.5            & \bf -0.40            & $ \bf 1.01^{+0.04}_{-0.03}$ & $\bf 0.88^{+0.06}_{-0.17}$ & \bf 66.5/50         & \bf 7.2            & \bf -0.35            \\

       & 28 & 178 & $1.00^{+0.11}_{-0.06}$ & $0.83^{+0.07}_{-0.07}$ & 66.0/50         & 2.8            & -0.44            & $1.01^{+0.04}_{-0.04}$ & $0.84^{+0.06}_{-0.06}$ & 63.7/50         & 10.1            & -0.13            \\
          & 29 & 179 & $1.00^{+0.07}_{-0.05}$ & $0.83^{+0.06}_{-0.07}$ & 75.6/50         & 4.7            & -0.48            & $1.02^{+0.05}_{-0.04}$ & $0.83^{+0.08}_{-0.06}$ & 54.5/50         & 8.1            & -0.80            \\
          & 30 & 180 & $1.01^{+0.08}_{-0.05}$ & $0.82^{+0.08}_{-0.08}$ & 89.7/50         & 3.0            & -0.68            & $1.02^{+0.04}_{-0.04}$ & $0.85^{+0.12}_{-0.12}$ & 62.4/50         & 7.8            & -0.86            \\
          & 31 & 181 & $1.00^{+0.06}_{-0.05}$ & $0.83^{+0.08}_{-0.08}$ & 75.3/50         & 3.5            & -0.35            & $1.00^{+0.04}_{-0.04}$ & $0.92^{+0.06}_{-0.27}$ & 62.7/50         & 5.5            & -0.69            \\

\hline
\hline       
\end{tabular}
\end{table}

\subsection{Comparison with previous BAO measurements}

\begin{figure}[t]
\centering
\includegraphics[width=0.5\textwidth]{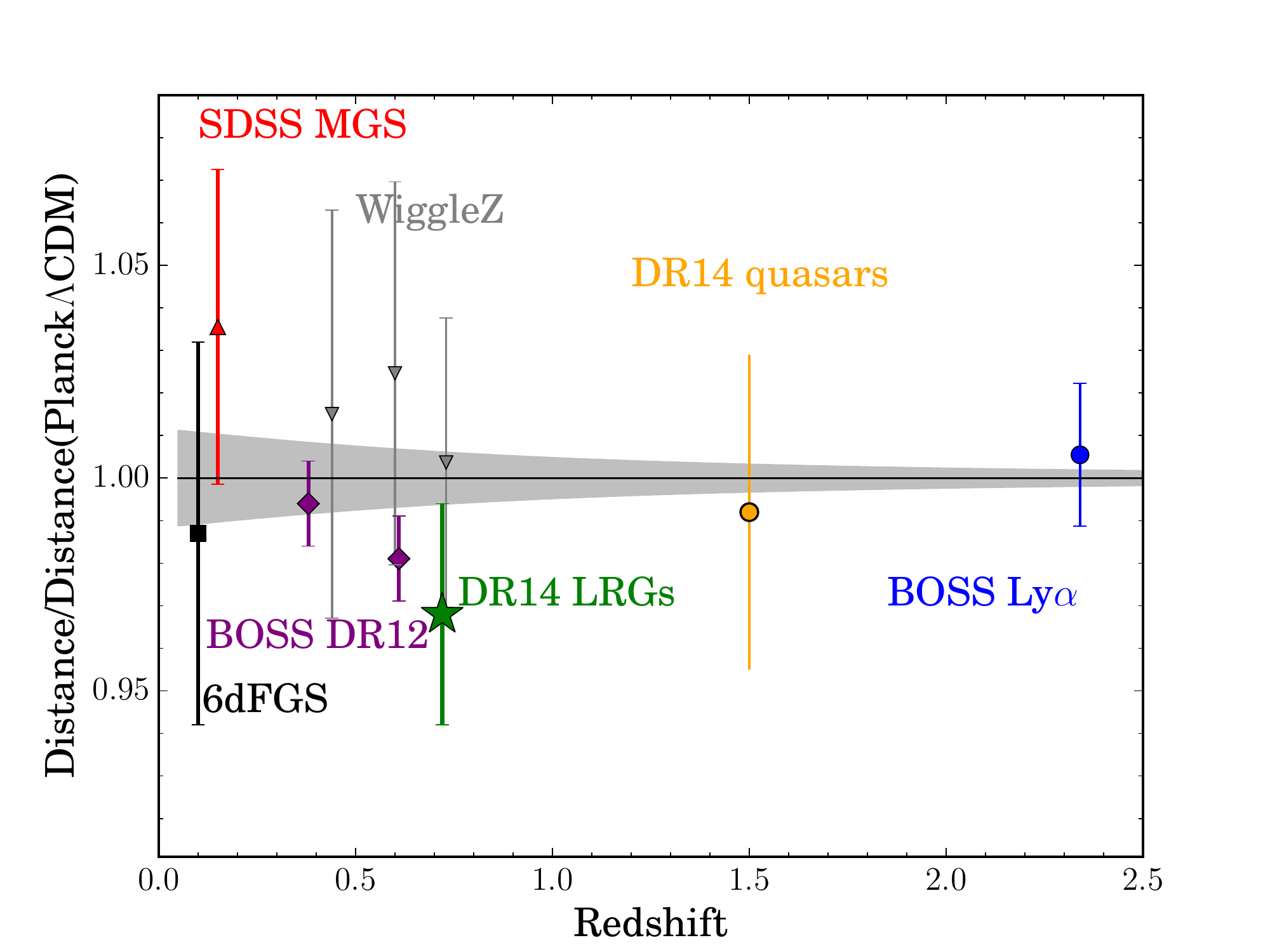}
\caption{Current isotropic BAO measurements as a function of 
redshift compared to the 
prediction given by the best-fit cosmological parameters of Planck 
TT+TE+EE+lowP \citep{Planck16XIII}. 
Our measurement is indicated by the green star labeled ``DR14 LRGs''. 
The other BAO measurements are: 
6dFGRS at $z=0.11$ \citep{Beutler11},
SDSS MGC at $z=0.15$ \citep{Ross15},
BOSS DR12 at $z=[0.38, 0.61]$ \citep{Alam17}, 
WiggleZ at $z=[0.44, 0.6, 0.73]$ \citep{Blake11},
eBOSS DR14 QSO sample at $z=1.52$ \citep{Ata17},
and BOSS DR12 Lyman-$\alpha$ sample at $z=2.3$ \citep{Bautista17, Bourboux17}. }
\label{fig:BAOrelPlanck}
\end{figure}

We summarize current distance measurements using BAO in 
Figure~\ref{fig:BAOrelPlanck}. The distances are normalized to the 
predictions using a Planck cosmology. Our measurement of 
the isotropic BAO scale at $z=0.72$ 
is consistent with that of Planck at about the 1$\sigma$ level.

Given that our eBOSS LRG sample was combined with the high-redshift tail 
of the CMASS sample in overlapping areas, the BAO measurements at 
$0.50<z<0.75$ from CMASS and ours, at $0.6<z<1.0$, are correlated. 
This correlation can be estimated by assuming that the covariance is 
proportional to the effective overlap volume between the two surveys. 
Using the CMASS $\bar{n}(z)$ over the effective area of eBOSS 
(overlapping area) covering 1844~deg$^2$, and 
computing Eq.~\ref{eq:veff} over $0.6 < z < 0.75$ (overlapping 
redshift range), 
we obtain an effective overlap volume of 
$V_{\rm eff}[{\rm CMASS \cap eBOSS}] = 0.31$~Gpc$^3$. 
Therefore, we estimate the correlation coefficient 
between the two measurements to be
\begin{equation}
\rho = \frac{V_{\rm eff}({\rm CMASS \cap eBOSS})}
{\sqrt{V_{\rm eff}({\rm CMASS}) V_{\rm eff}({\rm eBOSS})}} = 
\frac{0.31}{\sqrt{4.1 \times 0.9}} = 0.16.
\end{equation}
We leave more realistic calculations of this correlation using 
correlated mock catalogs (as in, e.g., \citealt{Beutler16}) for future work.

Forecasts in \citet{Zhao16} predict 1\% precision on 
isotropic BAO with 7000 deg$^2$ for the final eBOSS LRG sample 
(when combined with the high-redshift tail of CMASS). 
For the current footprint with $A_{\rm eff} = 1844$~deg$^2$ 
the forecast scales to a 1.95\% BAO measurement assuming that 
error is proportional to square-root of the effective volume. 
Our isotropic BAO measurement with a 2.6\% error is slightly larger
than this forecast. This might be caused by holes in the footprint due 
to plates still not observed and to the various masks applied to our 
sample. 
These effects increase the size of boundaries of the survey and might 
increase errors relative to forecasts which consider uniform volumes.
The larger error of our measurement compared to the forecast might 
also be due to statistical fluctuations, since the distribution
of estimated errors has a large dispersion, as observed 
with mock catalogs in Fig.~\ref{fig:hists_iso}. 
Previous BAO measurements (e.g. \citealt{Alam17, Ata17}) 
typically also report estimated errors that are larger 
than predictions.

\section{Conclusion}

We present the first BAO measurement using luminous red galaxies 
from the first two years of data taken in the eBOSS survey.
The total area observed, weighted by the fiber completenes, is 1844~deg$^2$, 
yielding an effective volume of 0.9~Gpc$^3$ over $0.6<z<1.0$ when combining
the eBOSS LRG sample with the CMASS $z>0.6$ galaxies over the eBOSS footprint.    
We obtain a 2.6\% spherically averaged distance 
measurement after reconstruction at $z_{\rm eff} = 0.72$ 
that is consistent at 1$\sigma$ level with the predictions of 
the $\Lambda$CDM model assuming a Planck best-fit cosmology.

In this analysis we introduce a novel technique to account for 
redshift failures, while also propagating photometric systematics to the 
random catalog. This technique yields unbiased measurements of the
correlation function, as tested on mock catalogs, and will be essential
for future analyses using the full-shape information such as redshift
space distortion studies.
 
When eBOSS will have finished its observing program, we expect that
 7000~deg$^2$ of area will have been observed spectroscopically, 
 representing a reduction
on errors of isotropic BAO measurements of a factor of 
$\sqrt{7000/1844} \sim 2$ (assuming errors scale with the square root  of 
the area). 

The new software used to produce catalogs,  compute model for failures, 
 fit BAO peak, and apply reconstruction are all implemented 
in Python and available at 
\url{github.com/julianbautista/eboss_clustering}. 

Upcoming surveys will significantly improve upon our results; 
the DESI survey will observe LRG spectra with similar depths and redshift 
ranges than eBOSS. 
We expect that the framework presented here should be applicable 
for DESI clustering measurements using both LRGs and ELGs, 
where sub-percent errors on BAO are 
expected.

\acknowledgments

This paper represents an effort by both the SDSS-III and SDSS-IV collaborations.
Funding for SDSS-III was provided by the Alfred
P. Sloan Foundation, the Participating Institutions, the
National Science Foundation, and the U.S. Department
of Energy Office of Science.
Funding for the Sloan Digital Sky Survey IV has been provided by
the Alfred P. Sloan Foundation, the U.S. Department of Energy Office of
Science, and the Participating Institutions. SDSS-IV acknowledges
support and resources from the Center for High-Performance Computing at
the University of Utah. The SDSS web site is www.sdss.org.

SDSS-IV is managed by the Astrophysical Research Consortium for the
Participating Institutions of the SDSS Collaboration including the
Brazilian Participation Group, the Carnegie Institution for Science,
Carnegie Mellon University, the Chilean Participation Group,
the French Participation Group, Harvard-Smithsonian Center for Astrophysics,
Instituto de Astrof\'isica de Canarias, The Johns Hopkins University,
Kavli Institute for the Physics and Mathematics of the Universe (IPMU) /
University of Tokyo, Lawrence Berkeley National Laboratory,
Leibniz Institut f\"ur Astrophysik Potsdam (AIP),
Max-Planck-Institut f\"ur Astronomie (MPIA Heidelberg),
Max-Planck-Institut f\"ur Astrophysik (MPA Garching),
Max-Planck-Institut f\"ur Extraterrestrische Physik (MPE),
National Astronomical Observatory of China, New Mexico State University,
New York University, University of Notre Dame,
Observat\'ario Nacional / MCTI, The Ohio State University,
Pennsylvania State University, Shanghai Astronomical Observatory,
United Kingdom Participation Group,
Universidad Nacional Aut\'onoma de M\'exico, University of Arizona,
University of Colorado Boulder, University of Portsmouth,
University of Utah, University of Virginia, University of Washington, University of Wisconsin,
Vanderbilt University, and Yale University.

This paper includes targets derived from the images of
the Wide-Field Infrared Survey Explorer, which is a
joint project of the University of California, Los Angeles,
and the Jet Propulsion Laboratory/California Institute
of Technology, funded by the National Aeronautics and
Space Administration.

The support and resources from the Center for 
High Performance Computing at the University of Utah are 
gratefully acknowledged.

MV is partially supported by Programa de Apoyo a Proyectos de 
Investigaci\'on e Innovaci\'on Tecnol\'ogica (PAPITT) No  IA102516, 
Proyecto Conacyt Fronteras No 281 and from Proyecto LANCAD-UNAM-DGTIC-319.

WJP \& E-MM acknowledge support from the UK Science and Technology 
Facilities Council grant ST/N000668/1. 
WJP also acknowledges support from the European Research Council 
through the Darksurvey grant 614030  and the UK Space Agency 
grant ST/N00180X/1.

\input{references_lrg.tex}
\end{document}

%% file: authors.tex
\author{Julian~E.~Bautista\altaffilmark{1},
Mariana~Vargas-Maga\~na\altaffilmark{2},
Kyle~S.~Dawson\altaffilmark{1},
Will~J.~Percival\altaffilmark{3},
Jonathan~Brinkmann\altaffilmark{4},
Joel~Brownstein\altaffilmark{1},
Benjamin~Camacho\altaffilmark{2},
Johan~Comparat\altaffilmark{5},
Hector~Gil-Mar{\'i}n\altaffilmark{6,7},
Eva-Maria~Mueller\altaffilmark{3},
Jeffrey~A.~Newman\altaffilmark{8},
Abhishek~Prakash\altaffilmark{9}
Ashley~J.~Ross\altaffilmark{10},
Donald~P.~Schneider\altaffilmark{11,12},
Hee-Jong~Seo\altaffilmark{13}
Jeremy~Tinker\altaffilmark{14},
Rita~Tojeiro\altaffilmark{15},
Zhongzu~Zhai\altaffilmark{14},
Gong-Bo~Zhao\altaffilmark{16}
}

\altaffiltext{1}{
Department of Physics and Astronomy, 
University of Utah, Salt Lake City, UT 84112, USA.
}

\altaffiltext{2}{
Instituto de Fis\'ica,
Universidad Nacional Autonoma de Mexico, Apdo.
Postal 20-364, 01000,Mexico, D.F.
}

\altaffiltext{3}{
Institute of Cosmology \& Gravitation, Dennis Sciama Building, University of Portsmouth, Portsmouth, PO1 3FX, UK.
}

\altaffiltext{4}{
Apache Point Observatory, P.O. Box 59, Sunspot, NM 88349, USA.
}

\altaffiltext{5}{
Max-Planck-Institut f\"ur Extraterrestrische Physik,
Giessenbachstra{\ss}e,
85748 Garching, Germany.
}

\altaffiltext{6}{
Sorbonne Universit\'{e}s, 
Institut Lagrange de Paris (ILP), 
98 bis Boulevard Arago, 75014 Paris, France
}

\altaffiltext{7}{
LPNHE, CNRS/IN2P3, Universit\'e Pierre et Marie
Curie Paris 6, Universit\'e ́ Denis Diderot Paris 7,
4 place Jussieu, 75252 Paris CEDEX, France
}

\altaffiltext{8}{
Department of Physics and Astronomy and PITT PACC,
University of Pittsburgh, Pittsburgh, PA 15260, USA.
}

\altaffiltext{9}{
Department of Astronomy, California Institute of Technology, Pasadena,
CA 91125, USA.
}

\altaffiltext{10}{
Center for Cosmology and Astro-Particle Physics,
Ohio State University, Columbus, OH 43210
}

\altaffiltext{11}{
Department of Astronomy and Astrophysics, 525 Davey Laboratory,
The Pennsylvania State University, University Park, PA 16802, USA.
}

\altaffiltext{12}{
Institute for Gravitation and the Cosmos,
The Pennsylvania State University, University Park, PA 16802, USA.
}

\altaffiltext{13}{
Department of Physics and Astronomy,
Ohio University,
251B Clippinger Labs, Athens, OH 45701
}

\altaffiltext{14}{
Center for Cosmology and Particle Physics,
Department of Physics, New York University,
4 Washington Place, New York, NY 10003, USA.
}

\altaffiltext{15}{
School of Physics and Astronomy, 
University of St Andrews, 
St Andrews, KY16 9SS, UK
}

\altaffiltext{16}{
National Astronomy Observatories, 
Chinese Academy of Science, 
Beijing, 100012, P. R. China.
}

\email{bautista@astro.utah.edu}